\documentclass[twocolumn]{aastex702}
\usepackage{amsmath,amssymb}

\newcommand{\per}{\ensuremath{^{-1}}}
\newcommand{\persq}{\ensuremath{^{-2}}}
\newcommand{\Ha}{H\ensuremath{\alpha}}
\newcommand{\Hb}{H\ensuremath{\beta}}
\newcommand{\mbh}{\ensuremath{M_{\textrm{BH}}}}
\newcommand{\betauv}{\ensuremath{\beta_{\textrm{UV}}}}
\newcommand{\betaopt}{\ensuremath{\beta_{\textrm{opt}}}}

\newcommand{\jwst}{\textit{JWST}}
\newcommand{\bhmf}{black hole mass function}
\newcommand{\lbol}{\ensuremath{L_{\textrm{bol}}}}

\shorttitle{The Black Hole Mass Function at $7<z<9.5$}
\shortauthors{Taylor et al.}

\begin{document}

\title{Little Red Dots with Broad H$\beta$ Lines Reveal the Black Hole Mass Function at $7<z<9.5$}

%% Lead Authors

\correspondingauthor{Anthony~J.~Taylor}
\email{anthony.taylor@austin.utexas.edu}

\author[0000-0003-1282-7454]{Anthony J. Taylor}
\email{anthony.taylor@austin.utexas.edu}
\affiliation{Department of Astronomy, The University of Texas at Austin, Austin, TX, USA}
\affiliation{Cosmic Frontier Center, The University of Texas at Austin, Austin, TX, USA}

\author[0000-0001-8519-1130]{Steven L. Finkelstein}
\email{stevenf@astro.as.utexas.edu}
\affiliation{Department of Astronomy, The University of Texas at Austin, Austin, TX, USA}
\affiliation{Cosmic Frontier Center, The University of Texas at Austin, Austin, TX, USA}

\author[0000-0003-3596-8794]{Hollis B. Akins}
\email{hollis.akins@utexas.edu}
\affiliation{Department of Astronomy, The University of Texas at Austin, Austin, TX, USA}

\author[0000-0002-5588-9156]{Vasily Kokorev}
\email{vkokorev@utexas.edu}
\affiliation{Department of Astronomy, The University of Texas at Austin, Austin, TX, USA}
\affiliation{Cosmic Frontier Center, The University of Texas at Austin, Austin, TX, USA}

%% Key Data / Simulation Architects/PIs

\author[0000-0002-7959-8783]{Pablo Arrabal Haro}
\email{pablo.arrabalharo@nasa.gov}
\affiliation{Center for Space Sciences and Technology, UMBC, 5523 Research Park Dr, Baltimore, MD 21228 USA}
\affiliation{Astrophysics Science Division, NASA Goddard Space Flight Center, 8800 Greenbelt Rd, Greenbelt, MD 20771, USA}

\author[0000-0002-1528-5091]{Sophie Koudmani}
\email{s.koudmani@herts.ac.uk}
\affiliation{Centre for Astrophysics Research, Department of Physics, Astronomy and Mathematics, University of Hertfordshire, College Lane, Hatfield, AL10 9AB, UK}
\affiliation{Kavli Institute for Cosmology, Cambridge, University of Cambridge, Madingley Road, Cambridge, CB3 0HA, UK}

\author[0000-0002-0930-6466]{Caitlin M. Casey}
\email{cmcasey@ucsb.edu}
\affiliation{Department of Physics, University of California, Santa Barbara, CA 93106, USA}
\affiliation{Cosmic Dawn Center (DAWN), Niels Bohr Institute, University of Copenhagen, Jagtvej 128, K{\o}benhavn N, DK-2200, Denmark}

\author[0000-0001-5414-5131]{Mark Dickinson}
\email{mark.dickinson@noirlab.edu}
\affiliation{NSF's National Optical-Infrared Astronomy Research Laboratory, 950 N. Cherry Ave., Tucson, AZ 85719, USA}

\author[0000-0002-6038-5016]{Junehyoung Jeon}
\email{junehyoungjeon@utexas.edu}
\affiliation{Department of Astronomy, The University of Texas at Austin, Austin, TX, USA}
\affiliation{Cosmic Frontier Center, The University of Texas at Austin, Austin, TX, USA}

%% Co-authors

\author[0000-0001-5384-3616]{Madisyn Brooks}
\email{madisyn.brooks@uconn.edu}
\altaffiliation{NSF Graduate Research Fellow}
\affiliation{Department of Physics, 196 Auditorium Road, Unit 3046, University of Connecticut, Storrs, CT 06269, USA}

\author[0000-0003-0212-2979]{Volker Bromm}
\email{bromm@astro.as.utexas.edu}
\affiliation{Department of Astronomy, The University of Texas at Austin, Austin, TX, USA}
\affiliation{Cosmic Frontier Center, The University of Texas at Austin, Austin, TX, USA}

\author[0000-0003-1949-7638]{Christopher J. Conselice}
\email{conselice@manchester.ac.uk}
\affiliation{Jodrell Bank Centre for Astrophysics, University of Manchester, Oxford Road, Manchester M13 9PL, UK}

\author[0000-0001-8047-8351]{Kelcey Davis}
\email{kelcey.davis@uconn.edu}
\altaffiliation{NSF Graduate Research Fellow}
\affiliation{Department of Physics, 196 Auditorium Road, Unit 3046, University of Connecticut, Storrs, CT 06269, USA}

\author[0000-0002-7622-0208]{Callum T. Donnan}
\email{callum.donnan@noirlab.edu}
\affiliation{NSF's National Optical-Infrared Astronomy Research Laboratory, 950 N. Cherry Ave., Tucson, AZ 85719, USA}

\author[0000-0002-9382-9832]{Andreas L. Faisst}
\affiliation{IPAC, California Institute of Technology, 1200 E. California Blvd. Pasadena, CA 91125, USA}
\email{afaisst@caltech.edu}

\author[0000-0001-7201-5066]{Seiji Fujimoto}
\email{fujimoto@utexas.edu}
\affiliation{Department of Astronomy, The University of Texas at Austin, Austin, TX, USA}

\author[orcid=0000-0001-6278-032X]{Lukas J. Furtak}
\affiliation{Department of Astronomy, The University of Texas at Austin, Austin, TX, USA}
\affiliation{Cosmic Frontier Center, The University of Texas at Austin, Austin, TX, USA}
\email{furtak@utexas.edu}

\author[0000-0002-7831-8751]{Mauro Giavalisco}
\email{mauro@umass.edu}
\affiliation{University of Massachusetts Amherst, 710 North Pleasant Street, Amherst, MA 01003-9305, USA}

\author[0000-0003-4242-8606]{Ansh R. Gupta}
\email{anshrg@utexas.edu}
\altaffiliation{NSF Graduate Research Fellow}
\affiliation{Department of Astronomy, The University of Texas at Austin, Austin, TX, USA}
\affiliation{Cosmic Frontier Center, The University of Texas at Austin, Austin, TX, USA}

\author[0000-0002-3301-3321]{Michaela Hirschmann}
\email{michaela.hirschmann@epfl.ch}
\affiliation{Institute of Physics, Laboratory of Galaxy Evolution, Ecole Polytechnique Federale de Lausanne (EPFL), Observatoire de Sauverny, 1290 Versoix, Switzerland}

\author[0000-0003-4512-8705]{Tiger Yu-Yang Hsiao}
\email{tiger.hsiao@utexas.edu}
\affiliation{Department of Astronomy, The University of Texas at Austin, Austin, TX, USA}
\affiliation{Cosmic Frontier Center, The University of Texas at Austin, Austin, TX, USA}

\author[0000-0002-8360-3880]{Dale D. Kocevski}
\email{dkocevsk@colby.edu}
\affiliation{Department of Physics and Astronomy, Colby College, Waterville, ME 04901, USA}

\author[0000-0002-6610-2048]{Anton M. Koekemoer}
\email{koekemoer@stsci.edu}
\affiliation{Space Telescope Science Institute, 3700 San Martin Drive, Baltimore, MD 21218, USA}

\author[0000-0002-9393-6507]{Gene C. K. Leung}
\email{gckleung@mit.edu}
\affiliation{MIT Kavli Institute for Astrophysics and Space Research, 77 Massachusetts Ave., Cambridge, MA 02139, USA}

\author[0000-0003-1581-7825]{Ray A. Lucas}
\email{lucas@stsci.edu}
\affiliation{Space Telescope Science Institute, 3700 San Martin Drive, Baltimore, MD 21218, USA}

\author[0000-0002-8951-4408]{Lorenzo Napolitano}
\email{lorenzo.napolitano@inaf.it}
\affiliation{INAF - Osservatorio Astronomico di Roma, via Frascati 33, 00078, Monte Porzio Catone, Italy}

\author[0000-0001-7503-8482]{Casey Papovich}
\email{papovich@tamu.edu}
\affiliation{Department of Physics and Astronomy, Texas A\&M University, College Station, TX, 77843-4242 USA}
\affiliation{George P.\ and Cynthia Woods Mitchell Institute for Fundamental Physics and Astronomy, Texas A\&M University, College Station, TX, 77843-4242 USA}

\author[0000-0001-8940-6768]{Laura Pentericci}
\email{laura.pentericci@inaf.it}
\affiliation{INAF - Osservatorio Astronomico di Roma, via Frascati 33, 00078, Monte Porzio Catone, Italy}

\author[0000-0003-4528-5639]{Pablo G. P\'erez-Gonz\'alez}
\email{pgperez@cab.inta-csic.es}
\affiliation{Centro de Astrobiolog\'{\i}a (CAB), CSIC-INTA, Ctra. de Ajalvir km 4, Torrej\'on de Ardoz, E-28850, Madrid, Spain}

\author[0000-0002-6748-6821]{Rachel S. Somerville}
\email{rsomerville@flatironinstitute.org}
\affiliation{Center for Computational Astrophysics, Flatiron Institute, 162 5th Avenue, New York, NY, 10010, USA}

\author[0000-0003-4793-7880]{Fabian Walter}
\affiliation{Max Planck Institut f\"ur Astronomie, K\"onigstuhl 17, D-69117 Heidelberg, Germany}
\affiliation{California Institute of Technology, Pasadena, CA 91125, USA}
\email{walter@mpia.de}

\author[0000-0003-3466-035X]{{L. Y. Aaron} {Yung}}
\affiliation{Institute of Astronomy and Department of Physics, National Tsing Hua University, Hsinchu 30013, Taiwan}
\email{yung@phys.nthu.edu.tw}

\begin{abstract}
We present a systematic search for broad \Hb{} emission at $7<z<9.5$ in 21,770 \jwst{}/NIRSpec PRISM spectra drawn from four blank fields and six lensing clusters with photometry spanning 1942 arcmin$^2$. We find a sample of 20 sources with statistically significant broad-\Hb{} emission, including 13 new sources. Our entire sample has colors and morphologies consistent with little red dots (LRDs), compared to only 37\% of broad-line emitters at $3.5<z<6$. Interpreting the broad-lines as indicators of Broad-line Active Galactic Nuclei (BLAGN), we infer black hole masses of $4.2\times10^{6}-2.1\times10^{8}$\,$M_{\odot}$. We calculate completeness corrections and construct the \bhmf{} at $z\sim8$. The normalization of the \bhmf{} rises by $\approx1.0$~dex from $z\sim8-5$, consistent with average black hole growth at $f_{\textrm{Edd}}\approx0.1$. Comparing with cosmological simulations, the $z\sim8$ \bhmf{} requires heavy ($10^5$\,$M_{\odot}$) seeds or super-Eddington accretion onto light or intermediate mass seeds to match its normalization.  Finally, we infer bolometric luminosities for the sample and construct a bolometric luminosity function. We integrate over this luminosity function to derive a black hole accretion density (BHAD) of $4.8^{+5.8}_{-2.6}\times10^{-7}~M_{\odot}~\textrm{yr}^{-1}~\textrm{Mpc}^{-3}$, showing broad agreement with previous photometric work, but disagreement with previous X-ray models, reinforcing the idea that BLAGN are more common in the early universe than previously expected.

\end{abstract}

\section{Introduction} \label{sec:intro}

Understanding the evolution of Supermassive Black Holes (SMBHs) across cosmic time is a key component in developing a full understanding of both the evolution of the Universe and the formation of galaxies \citep[e.g.,][]{heckman14}. As most if not all local massive galaxies are known to host SMBHs at their centers, SMBHs and galaxy properties and evolution are intrinsically tied together. This co-evolution is most apparent when SMBHs actively accrete matter as Active Galactic Nuclei (AGN), which strongly radiate across the electromagnetic spectrum. This emission is thought to significantly influence the star formation and gas dynamics of the host galaxy \citep{kormendy_ho2013}. In broad-line AGN (BLAGN), the emergent radiation from the SMBH's accretion disk that is processed through the SMBH's high-density broad-line region is directly visible, and is detectable through Doppler broadening of permitted emission lines (e.g.\ Mg\,\textsc{ii}, Ly$\alpha$, \Ha{}, \Hb{}, Pa$\alpha$) and a lack of broadening in density-sensitive forbidden lines (e.g.\ [O\,\textsc{iii}]$\lambda\lambda$4959,5007). Due to this direct view of the emitting region, BLAGN are among the most luminous known objects in the universe \citep[e.g.][]{sanders88}. 

As a result of this high brightness and characteristic broad emission lines, BLAGN are straightforward to identify in spectroscopy, even at high redshifts. The limiting factor in identification with increasing redshift is spectral wavelength coverage, as the typically brightest emission line from BLAGN, \Ha{}, enters the near-infrared at $z\sim1$. Prior to the launch of the \textit{James Webb Space Telescope} \citep[\jwst{};][]{gardner06,gardner23}, the detection of BLAGN at $z>3$ was limited to only the brightest (and thus most massive $\gtrsim10^{8} M_{\odot}$) sources (typically quasars) using rest-UV emission features such as N\,\textsc{v}, C\,\textsc{iv}, or Mg\,\textsc{ii} \citep[e.g.][]{jiang16,yang21}. With the advent of \jwst{} and its unprecedented sensitivity to the near- and mid-infrared, the population of known BLAGN at $3<z<7$ (via \Ha{} observed at up to 5.3\,$\mu$m) has rapidly expanded to both fainter luminosities and lower masses \citep[e.g.][]{kocevski23,harikane23,kocevski25,taylor25a,juodzbalis25}.

These samples \citep[presently $\sim$tens of objects, e.g.][]{taylor25a,juodzbalis25,jones26} have enabled unprecedented studies of sub-quasar BLAGN at $z>3$. In parallel to these BLAGN populations, \jwst{} has also revealed a curious new population of sources, referred to as ``Little Red Dots'' (LRDs). The first of these objects with NIRSpec spectroscopy was presented in \cite{kocevski23}, and the name ``LRD'' was coined in \cite{matthee24}. LRDs were subsequently identified in surprising quantities in both photometric \citep[e.g.][]{akins25c,kocevski25,kokorev24,barro24b,PerezGonzalez24} and spectroscopic studies \citep[e.g.][]{greene24,taylor25a,juodzbalis25,kocevski25,hviding25,perezgonzalez26}. While the exact physical nature of LRDs is still hotly debated, it is commonly accepted that they exhibit blue rest-frame UV $\beta$ spectral slopes, distinctly red rest-frame optical $\beta$ slopes, are compact and unresolved in the rest-optical by \jwst/NIRCam, predominantly exhibit broad Balmer lines \citep[e.g.,][]{akins25c,kocevski23,kocevski25,barro24b,hviding25}. LRDs also tend to exhibit X-ray weakness, a lack of variability, and an apparent lack of dust emission \citep[e.g.,][]{yue24xray,maiolino25,kokubo25,burke26,PerezGonzalez24,leung24}.

While the nature of these sources remains uncertain, most prevailing theories suggest that LRDs host a powerful radiative central engine surrounded by a dense gas medium. Furthermore, while it is generally accepted that such a medium is capable of reproducing the red half of an LRD-like spectral energy distribution \citep[SED; e.g.][]{inayoshi25,naidu25,degraaff25,degraaff25c,taylor25c,pacucci26,jeon26}, within this framework, the identities of the central engines of LRDs are difficult to probe \citep{begelman26,gentile26}. Due primarily to the broad Balmer lines seen in most LRD spectra, BLAGN were obvious initial candidates for these central engines, with the broad Balmer lines attributed to AGN broad-line region kinematics \citep[e.g.][]{harikane23,kocevski23,kocevski25,maiolino24,juodzbalis25,taylor25a,barger26,piero26,brazzini26}. 

However, it was later suggested that the proposed dense enshrouding media could instead scatter much narrower Balmer lines to produce a similar profile to the observed broad-lines \citep{rusakov25,chang26,kokorev26}. In this model, the velocities in the AGN broad-line region would be much less than expected from the observed broad-lines, resulting in gross overestimates of black hole mass by $\sim1-2$~dex. Here bolometric-luminosity-based masses (assuming an Eddington ratio) may be more reliable \citep{perezgonzalez26,umeda26}. However, it is also important to note that the line structures predicted by this model have also been observed in more ordinary AGN without dense enshrouding media \citep{laor06}. 

An even more recent model instead posits that LRD engines are supermassive stars, and possible progenitors of globular clusters \citep{chisholm26}. Such supermassive stars could initially serve as the central engines, and would likely collapse into black holes in $\sim1$~Myr \citep{begelman10}, thereby feeding into one of the previously mentioned models. This model may also be supported by detections of Wolf-Rayet features in LRDs \citep{perezgonzalez26}. LRDs have also been modeled as ``Quasi-stars'' (QSs) \citep{begelman26}. QSs are closely related to the enshrouding model, as they feature super-Eddington accreting black holes at the center of proto-star-like gas envelopes. As such QS model is a variation of the overall DCBH framework, where the initial heavy seed is rapidly devouring the surrounding birth cocoon \citep[e.g.,][]{Pacucci2015,pacucci26,Becerra2018}. 

At the time of this work, all of these theories appear to remain plausible and competitive. While future studies may support or disprove these or future models, for the fiducial case in this work, we treat LRDs as BLAGN, whose broad Balmer lines directly reflect the energetics and kinematics of their broad-line regions before considering alternate interpretations. 

Samples of BLAGN at $z>3$ are instrumental in providing constraints on the high-redshift \bhmf{}---the number density of black holes as a function of black hole mass (\mbh{})---which is essential for understanding the evolution of SMBHs over cosmic time. The \bhmf{} may also provide insight into black hole growth and seeding \citep[e.g.,][]{Smith2019,Inayoshi2020}. Black hole seeds are typically classified as either light ($\lesssim 10^3 \, \rm M_\odot$), intermediate ($\sim10^3-10^4 \, \rm M_\odot$) or heavy ($\gtrsim 10^4 \, \rm M_\odot$). Light seeds are expected to form from the deaths of the first generation of massive stars \citep[likely Population~III stars, e.g.,][]{Madau2001,bromm04}. Intermediate seeds may form from rapid runaway mergers in dense star clusters, and may be common in high-redshift galaxies \citep[e.g.,][]{PZ_2002, Freitag_2006, Miller_2012, Katz_2015, Lupi_2016, Boekholt_2018,GonzalezKremer2021,dekel25a,dekel25b}.

Heavy seeds are instead proposed to be the result of the collapse of metal-free clouds at $z\gtrsim 10$, referred to as direct collapse black holes \citep[DCBHs; e.g.,][]{Loeb_Rasio_1994, Bromm_Loeb_2003, Lodato_Natarajan_2006}. In some cases, such as the ``naked'' SMBH in Abell~2744-QSO1 \citep{Maiolino_QSO1_2026}, the seeding may involve massive primordial black holes (PBHs), formed in the ultra-early Universe \citep[e.g.][]{Zhang_PBH2025}. While it is possible for light seeds to grow from early times to form the massive SMBHs seen locally via Eddington-limited accretion, the most massive SMBHs at $z \gtrsim 6$ have been found to be too large to have formed from Eddington-limited light-seeds \citep[e.g.][]{Inayoshi2020,fan23}. These early SMBHs must instead have formed via light seeds that undergo super-Eddington accretion or from intermediate to heavy initial seeds \citep[e.g.][]{Bromm_Loeb_2003,taylor25c,pacucci26}. 

In a previous study, \cite{taylor25a} constructed the \bhmf{} at $3.5<z<6$, using a population of broad-\Ha{} emitters (both LRDs and non-LRDs) and determined that any signatures of specific black hole seeding pathways (e.g. light Eddington-limited seeds, light super-Eddington seeds, or heavy seeds) are lost by $z\sim5$, as comparative models of all three cases were degenerate. There are two direct ways to make forward progress on this investigation. The first is to probe SMBHs to lower masses. This requires either deeper NIRSpec data, assistance from gravitational lensing, or spectral stacking of multiple objects \citep[e.g.][]{geris26,brooks26}. \cite{fei25} leveraged lensing to push the \bhmf{} down to $10^{5.5}$\,$M_{\odot}$, revealing tentative evidence for a rapid increase in the \bhmf{} at the lowest masses potentially indicative of heavy seeds \citep{jeon25}. The other way forward is to push to higher redshifts. In these early epochs, SMBHs have had less time to grow, and SMBH growth is more difficult, as gravitational potential wells are shallower, and stellar feedback may limit the supply of gas available for accretion. As such, the signatures of seeding may be more visible there. Here we aim to chart the latter path---constructing the \bhmf{} at $z\sim8$ by searching for broad-\Hb{} emitters. 

This work is organized as follows. In \S\ref{sec:observations} we summarize the photometry, spectroscopy, and lens models used in this analysis and discuss data reductions techniques when applicable. In \S\ref{sec:analysis}, we discuss our redshift finding methods and our broad-\Hb{} detection and fitting routine. In \S\ref{sec:sample}, we introduce our final sample of 20 $z\sim8$ BLAGN, and detail our completeness corrections in preparation for computing the \bhmf{}. In \S\ref{sec:results}, we present the \bhmf{}, the bolometric luminosity function and black hole accretion density, and the UV luminosity function of the sample. We then discuss and interpret these results with comparisons to models of black hole seeding and growth in \S\ref{sec:discussion}. Finally, in \S\ref{sec:summary}, we summarize the main conclusions of this work.

We assume $\Omega_m$=$0.3$, $\Omega_\Lambda$=$0.7$, and $H_0$=$70$~km~s$^{-1}$~Mpc$^{-1}$ throughout. All magnitudes are given in the AB magnitude system, where an AB magnitude is defined by $m_{AB}$=$-2.5\log f_\nu - 48.60$. Here, $f_\nu$ is the specific flux of the source in units of erg~cm$^{-2}$~s$^{-1}$~Hz$^{-1}$. We use `$\log$' to refer to the base-10 logarithm throughout.

\section{Observations}\label{sec:observations}

This study combines \jwst{} NIRSpec and NIRCam data in the Abell 2744, Abell 370, MACS~J0416, MACS~J0417, MACS~J1149, and MACS~J1423 gravitationally lensed fields as well as the COSMOS, EGS, GOODS-S, and UDS blank fields. NIRCam data were sourced from the CANUCS \citep[GO\#1208; DR1;][]{canucsdr1}, CEERS \citep[ERS\#1345;][]{finkelstein25}, COSMOS-Web \citep[GO\#1727;][]{cosmosweb23}, COSMOS-3D \citep[GO\#5893; e.g.][]{lin26}, JADES \citep[GTO/GO\#1180, 1286, 1287, 3215; GOODS-S;][]{jades26}, MIDIS \citep[GTO\#1283;][]{midis25}, NGDEEP \citep[GO\#2079;][]{ngdeep24}, PRIMER \citep[GO\#1837;][]{dunlop_2021_primer}, UNCOVER \citep[GO\#2561;][]{uncover}, and publicly available overlapping parallel programs. NIRSpec PRISM observations were compiled from the CANUCS \citep[GO\#1208; DR1;][]{canucsdr1}, CAPERS \citep[GO\#6368; e.g.][]{kokorev25}, COSMOS-DDT \citep[DDT\#6585;][]{coulter24}, CEERS \citep[ERS\#1345;][]{finkelstein25}, EMBER (GO\#7076; H.\,Akins et al.\,in prep.), JADES \citep[GTO/GO\#1180, 1286, 1287, 3215; GOODS-S;][]{jades26}, Mirage or Miracle \citep[MoM; GO\#5224; e.g.][]{naidu25}, RUBIES \citep[GO\#4233;][]{rubies}, UNCOVER \citep[GO\#2561;][]{uncover}, and ZENITH (GO\#7417; M.\,Franco in prep.) programs from observations taken up to 15 May 2026. Supplemental NIRSpec G395M grating observations were obtained from C3PO \citep[GO\#5943; e.g.][]{papovich26}, THRILS \citep[GO\#5507;][]{hutchison26}, and SPURS \citep[GO\#9214; e.g.][]{tang26,chen26a,chen26b}.

\subsection{Photometry}\label{sec:photometry}

Photometric reduction was performed on a field-by-field basis, using public imaging datasets when available. We source our GOODS-S imaging from JADES DR5 \citep{johnson26}. For COSMOS, EGS, and UDS, we use an internal reduction from the CAMPFIRE pipeline\footnote{\url{https://github.com/hollisakins/campfire}} (full details will be described in H.~Akins~et~al., in prep.). For MACS~J0416, MACS~J0417, MACS~J1149, MACS~J1423, Abell~2744, and Abell~370, we use custom reductions from the \texttt{grizli} pipeline \citep{grizli}. For the full details, see V.~Kokorev~et~al., in prep.

We used the photometry catalogs from the UNICORN project (S. Finkelstein et al., in prep).  These consist of multi-wavelength photometry, using the NIRCam long-wavelength imaging as the detection image.  Care is taken to measure robust colors, total fluxes, and empirical uncertainties using well-tested criteria stemming from the CEERS survey.  All catalogs come with photometric redshift measurements calculated with Lazy.jl \citep{lazy_jl}, using a set of templates optimized to yield excellent redshift agreement with existing NIRSpec spectroscopy.

\subsection{Spectroscopy}\label{sec:spectroscopy}

We re-reduce all collected PRISM spectra with the same procedure to ensure a uniform spectral sample. We use the \jwst{} Calibration Pipeline\footnote{\url{https://github.com/spacetelescope/jwst}} \citep{bushouse23} version 1.20.2 and CRDS version \texttt{1464.pmap}. We primarily use the default configuration for the pipeline with a few exceptions. First, we enable the \texttt{clean\_flicker\_noise} step (using the ``median" method) in \texttt{calwebb\_detector1} to reduce the effects of 1/f noise. We also use a modified flat-field file in the \texttt{calwebb\_spec2} stage \citep[see][for details]{arrabalharo23}. Due to the large diversity of programs and corresponding observing strategies used in this work, we combine individual NIRSpec nods using the \texttt{calwebb\_spec3} routine on a visit-by-visit basis, marginalizing over any dithers or MSA reconfigurations. We use an optimal extraction \citep{horne86} to best extract the target signal from the 2D spectra to produce 1D spectra. We calibrate these 1D spectra to \jwst{}/NIRCam photometry (see \S\ref{sec:photometry}) by computing the flux in the spectrum within each of the F150W, F200W, F277W, F356W, and F444W NIRCam filters. We fit a third-order Chebyshev polynomial to the residuals between the fluxes measured from the spectrum and from aperture photometry. We then apply this fitted correction curve to the spectra to mitigate the effects of NIRSpec path-loss and related corrections applied in the \jwst{} Calibration Pipeline that are not tied to NIRCam data. Objects that are not covered by \jwst{}/NIRCam imaging are removed from the sample and not considered in subsequent analysis. We correct all spectra for the effects of gravitational lensing when appropriate (see \S\ref{sec:lens}).

For objects that were observed in multiple visits (especially across different observing programs), we combine their calibrated and optimally extracted 1D spectra by interpolating all spectra to the wavelength grid of the highest exposure time spectrum using \texttt{spectres} \citep{spectres}, and then stacking with an inverse variance weighting. This approach offers a balance between computational complexity and data quality, especially for programs that employed complicated observing strategies. We exclude exposures affected by MSA shorts (e.g., part of UNCOVER/GO\#2561 observation 2, visit 1) to ensure consistent data quality in the final 1D stacks. The resulting set of 21,770 objects with PRISM spectra span effective exposure times from $0.26-31.6$~hours.

\subsection{Lens Models}\label{sec:lens}

For the Abell 2744, MACS~J0416, MACS~J0417, MACS~J1149, MACS~J1423, and Abell 370 fields, we source gravitational lens models provided by the UNCOVER and CANUCS programs \citep{furtak23, canucsdr1}. We use the highest resolution models available within these data releases. 

\section{Data Analysis}\label{sec:analysis}

\subsection{Redshift Determination}\label{sec:zspec}

To search for broad \Hb{} emission at $7<z<9.5$, we first require robust spectroscopic redshifts (spec-zs) for all objects. While JADES and UNCOVER provide official catalogs of vetted redshifts \citep{curtislake26,price25} and we use internal vetted CAPERS and CEERS redshift catalogs, such products are not publicly available for CANUCS, COSMOS-DDT, EMBER, MoM, RUBIES, or ZENITH. We therefore develop a custom NIRSpec/PRISM spectroscopic redshift fitting code inspired in part by \texttt{msaexp} \citep{msaexp}. Briefly, our code uses a combination of continuum templates derived from \texttt{BAGPIPES} \citep{carnall18}, individual emission line templates, broad-line templates for \Ha{}, \Hb{}, and Pa$\alpha$, simple blackbody templates, and a modified blackbody template derived from the best-fit modified blackbody to ``The Cliff'' \citep{degraaff25} as fit in \cite{degraaff25b}. We pre-compute these templates on a fine redshift grid ($0<z<20$, $\Delta z=0.01$), convolving each with the PRISM line-spread function (LSF) curve. Next, for a given PRISM spectrum, we fit this collection of templates using non-negative least squares for all redshifts in the redshift grid. At each grid point, we compute an optimized fit and a corresponding $\chi^2$. We then select the redshift solution with the minimal $\chi^2$ as the best redshift for the object. 

Using the CAPERS human-vetted redshift sample as a testing set, this program returns an accurate ($|z_{fit}-z_{vetted}|/(1+z_{vetted})<0.05$) redshift for 97.2\% (4368/4491) of the test objects across all redshifts. For objects with vetted redshifts $z>7$, the program performs even better with a 98.6\% (142/144) accuracy. As such, this program reliably identifies $z>7$ objects---the targeted sample for this work---primarily through strong [O\,\textsc{iii}]$\lambda\lambda4959,5007$ lines and strong Ly$\alpha$ breaks. We use this program to measure spectroscopic redshifts for CANUCS, COSMOS-DDT, EMBER, MoM, RUBIES, and ZENITH. We visually inspect and verify that all objects fit at $z>7$ are indeed $z>7$ objects (based on a clear \Hb{} line and [O\,\textsc{iii}] doublet) to ensure a pure sample. This produces a set of 546 unique spec-z confirmed objects at $7<z<9.5$. When considering gravitationally lensed sources with multiple images, we restrict our analysis to the brightest image to avoid double-counting the source. 

\subsection{Broad-line Fitting and Statistical Testing} \label{sec:fitting}

\subsubsection{Initial PRISM FWHM Calibration} \label{sec:LSF1}

Systematically detecting and fitting broad-lines in PRISM spectra presents significant challenges over similar procedures in \jwst{}/NIRSpec medium or high-resolution gratings. Firstly, the spectral resolution of the PRISM disperser ($R\sim100$) is far lower than the medium ($R\sim1000$) or high-resolution gratings ($R\sim2700$). As such, distinguishing between a true kinematically broadened line (with an intrinsic full width at half maximum [FWHM] $\gtrsim 1000$\,km\,s\per{} and an observed (LSF-convolved) FWHM~$\gtrsim 3160$\,km\,s\per{}) and an unresolved line with an observed FWHM~$\sim 3000$\,km\,s\per{} (both derived simply from an assumed FWHM~$\sim 3000$\,km\,s\per{} LSF at $R\sim100$) is impractical. However, the spectral resolution curve of the PRISM provided by the \jwst{} documentation assumes a fully illuminated spectral slit. In reality, many NIRSpec targets are more compact than the $0\farcs2$ width of an MSA shutter, and thus exhibit a higher spectral resolution and a narrower LSF than the fully illuminated shutter curves might suggest. 

We address these challenges with an empirical calibration. Firstly, the PRISM's spectral resolution varies strongly with wavelength, such that even for a fully illuminated slit, the PRISM reaches a spectral resolution of $R\sim170-250$ for \Hb{} at $7<z<9.5$, corresponding to LSF FWHM of $1760-1200$~km~s\per. Such LSFs still allow for the detection of kinematic $1000$\,km\,s\per{} broad-lines, as the combination of the kinematic broad-line and the LSF will result in observed FWHMs of $2020-1600$\,km\,s\per{} that can be statistically differentiated from narrow lines that are dominated by the LSF. The LSF's dominance in determining the observed line widths of narrow lines allows for an empirical calibration of the source's effective LSF curve by fitting the [O\,\textsc{iii}]$\lambda\lambda4949,5007$ doublet. As the lines in this doublet should be kinematically very narrow ($\ll 500$~km~s\per{}) when compared to the LSF, the observed width of these lines can be used to empirically determine the LSF on a spectrum-by-spectrum basis. 

\subsubsection{Broad-line Models and Fitting} \label{sec:fit_model}

As such, we define a nine-parameter model capable of testing for broad \Hb{} emission in NIRSpec PRISM spectra at $7<z<9.5$. These parameters are: (1) the redshift of the source, (2) the continuum level near \Hb{}, (3) the continuum slope near \Hb{}, (4) the amplitude of the [O\,\textsc{iii}]$\lambda5007$ line, (5) the amplitude of the narrow \Hb{} line, (6) the amplitude of the broad \Hb{} line, (7) the kinematic FWHM of the narrow \Hb{} line; restricted to $<500$\,km\,s\per{}, (8) the kinematic FWHM of the broad \Hb{} line; restricted to $>700$\,km\,s\per{}, and (9) the FWHM of the LSF; restricted to between 0.5 ($\sim$1 pixel) and 1.0 times the fully illuminated slit curve. 

We model the [O\,\textsc{iii}]$\lambda\lambda4959,5007$ doublet as a pair of Gaussians with FWHMs determined entirely by the LSF, and a fixed 2.98:1 line flux ratio \citep{osterbrock06}. We model the \Hb{} emission line with two Gaussian components: one broad and one narrow, each with independent kinematic FWHMs (that are added in quadrature with the LSF to produce observed FWHMs) and line amplitudes. As the PRISM spectral resolution is insufficient to model a velocity offset between these components or robustly detect absorption on the \Hb{} line \citep[e.g.][]{inayoshi25}, we enforce a common line center for the broad and narrow components and do not model any narrow absorption. Finally, we use a simple linear model for the continuum with a slope and an offset level. As a standard of comparison, we also define a narrow-line-only model for \Hb{}, in which the broad Gaussian component of the \Hb{} line is excluded. 

We fit both of these models (broad+narrow-line and narrow-line-only) to all 546 sources at $7<z<9.5$ with confirmed redshifts using a Markov Chain Monte Carlo \citep[\texttt{emcee;}][]{emcee}. In order to qualify as a secure broad-line detection, we require that the difference in the Bayesian Information Criteria (BIC) of the fitted models is greater than six (commonly interpreted as strong evidence) in favor of the model which includes a broad-\Hb{} component. We further require that the signal-to-noise (S/N) on the broad-component of \Hb{} be greater than four to balance sample completeness with purity. These criteria yield a sample of 21 \Hb{} broad-line objects. 

\subsubsection{Outflow Detection} \label{sec:outflows}

We next verify that these broad \Hb{} lines are not dominated by outflows using \jwst{}/NIRSpec G395M data (available for 13 broad-line objects). Here we add two additional parameters to the broad-line model: (1) the amplitude of a broadened [O\,\textsc{iii}]$\lambda\lambda4959,5007$ outflow component (in all cases, much less than the narrow component amplitude, and thus undetected in the PRISM), and (2) the kinematic FWHM of the [O\,\textsc{iii}]$\lambda\lambda4959,5007$ outflow component. We combine these two parameters to add a second, broad Gaussian component to the [O\,\textsc{iii}]$\lambda\lambda4959,5007$ doublet. At the densities of an AGN broad-line region, the critical density of [O\,\textsc{iii}]$\lambda\lambda4959,5007$ is exceeded, suppressing any component of the line that would trace high-velocity AGN kinematics. However, outflow densities are much lower \citep[$n_e\sim10^2-10^4$~cm$^{-3}$; e.g.,][]{spence18,mingozzi19,baron19}, allowing the [O\,\textsc{iii}]$\lambda\lambda4959,5007$ doublet to serve as a tracer. For objects where the broad \Hb{} component is from a BLAGN, no significant broad component of [O\,\textsc{iii}]$\lambda\lambda4959,5007$ with a FWHM comparable to \Hb{} should be detected, while objects with a dominant outflow may show broadened [O\,\textsc{iii}]$\lambda\lambda4959,5007$, comparable in FWHM to the broad \Hb{}.

We fit both this outflow model and the previous \Hb{} broad-line model to the 13 broad-line objects that have G395M data (with stacked effective exposure times varying from 0.9--11.0~hours). For an object to have a significant outflow component, we require (1) the BIC is greater than six in favor of the outflow component and (2) the S/N of the outflow component is greater than 3. Three objects satisfy these criteria: J095948.35+022655.5, J141930.27+525251.0, and J142008.49+525326.4. We next examine the FWHMs of the outflow components to see if they match the FWHM of the broad component of \Hb{}. In objects J095948.35+022655.5 and J141930.27+525251.0, the outflow FWHMs are both $\lesssim1000$\,km\,s\per{}, while the broad component of \Hb{} FWHMs are both $\gtrsim3000$\,km\,s\per{}, indicating that the outflow component seen in [O\,\textsc{iii}]$\lambda\lambda4959,5007$ is kinematically inconsistent with the BLAGN signature exhibited in \Hb{}. However, the third object J142008.49+525326.4 \citep{larson23} exhibits an outflow FWHM of $\sim900$\,km~s\per{} and a broad component of \Hb{} FWHM of $\sim1400$\,km~s\per. This similar but inexact match in FWHM likely indicates that at least one component (J142008.49+525326.4 has been shown to exhibit a clumpy spatial structure; \citealt{larson23}) is a BLAGN, but J142008.49+525326.4 also hosts strong outflows that significantly contribute to the broad \Hb{} line. As performing the more detailed modeling necessary to measure the AGN kinematics in isolation from the outflow for a single object is beyond the scope of this population study, we exclude this object from the sample to produce a final sample of 20 objects.

\subsubsection{Further PRISM FWHM Calibration} \label{sec:fwhmcal}

As a by-product of the G395M outflow testing, more accurate broad-\Hb{} FWHMs can be measured for objects with sufficiently deep data. Here we require the S/N on the measured broad-\Hb{} flux to be greater than three. Four objects meet this criterion: J100003.90+020734.2, J095948.35+022655.5, J100120.74+021438.8, and J141930.27+525251.0. Despite allowing variably fit LSFs for both the PRISM and grating, and forward modeling the effects of the LSFs to derive the underlying intrinsic line FWHMs, the grating-measured FWHMs are---on average and accounting for the MCMC posterior uncertainties---only $81^{+9}_{-21}\%$ of the PRISM-measured FWHMs. Therefore, for the four objects with sufficiently deep G395M data, we adopt these more-accurate measurements, and we apply the $81^{+9}_{-21}\%$ correction factor to the other 16 objects in the sample (increasing the uncertainty on their FWHM measurements by rescaling each point of the MCMC FWHM posteriors by a random draw from the MCMC derived distribution of correction factors). Curiously, this FWHM effect is not exhibited in simulated NIRSpec PRISM spectra from \texttt{pandeia} \citep{pontoppidan16} used in our completeness simulations (see \S\ref{sec:detcomp}), thus it is likely that the assumed convolution of a Gaussian line with a Gaussian LSF may be inaccurate in real NIRSpec data. 

\begin{figure*}[t]
\includegraphics[width=\textwidth]{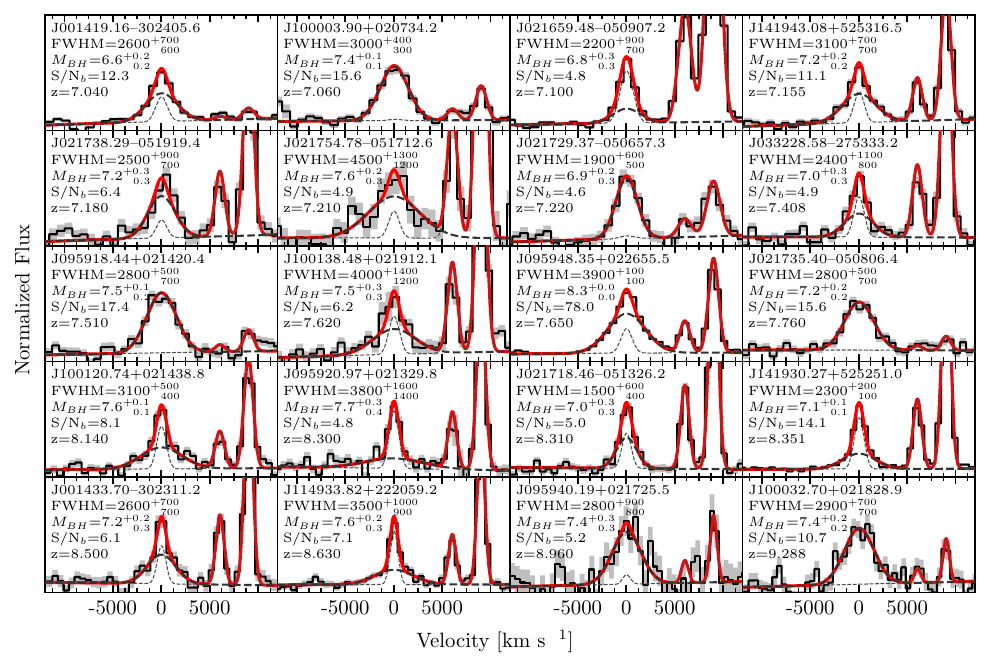}
\caption{A gallery of the 20 objects in our final sample of broad-\Hb{} emitters at $7<z<9.5$ ordered by ascending redshift.  We show the full broad-line and narrow-line profiles with red curves. We show the broad and narrow components for each object as dashed grey curves. We detail the main measurements for each object in Table~\ref{tab:sample}, and list the surveys and spectra for each object in Table~\ref{tab:apdx}. The final sample shows remarkable diversity in broad-line widths, broad/narrow line ratios, and [O\,\textsc{iii}]$\lambda\lambda4959,5007$ strengths.} \label{fig:sample}
\end{figure*}

\section{The Sample}\label{sec:sample}
We present the final sample of 20 $\Delta$BIC$>6$, broad-\Hb{} S/N$>4$, $z>7$ objects in Figure~\ref{fig:sample} and Table~\ref{tab:sample}. We detail the exact spectra and previous works (if any) that first presented the objects in Table~\ref{tab:apdx} (Appendix~\ref{apdx:table}). Of the 20 objects, 13 have not yet been presented in the literature as BLAGN. Of the seven previously presented objects, we present new deeper data for two objects (J095948.35+022655.5 and J141930.27+525251.0) through the inclusion of EMBER and CAPERS data, respectively. Our final sample fully recovers all seven known objects in the literature in the spectroscopic datasets searched in this work. At the time of writing, only three other broad-\Hb{} emitters at $z>7$ have been reported in the literature in G395M data and NIRCam WFSS observations \citep{xiao25,fei26}. 

\begin{deluxetable*}{rrrcccccccc}
\tablecaption{Measured Properties of BLAGN at $7<z<9.5$ \label{tab:sample}}
\tablehead{
Source ID & R.A. & Decl. & $z_{spec}$ & $F_{H \beta}$ & $F_{H \beta, n}$ & $F_{H \beta, b}$ & FWHM$_{H \beta, b}$ & EW$_{H \beta, b,0}$ & $\log_{10}(M_{BH})$ & $\mu$ \\
 & [deg] & [deg] &  & \multicolumn{3}{c}{[$10^{-18}$ erg s$^{-1}$ cm$^{-2}$]} & [km s$^{-1}$] & [\AA{}] & $[M_{\odot}]$} 
 \startdata
J001419.16-302405.6 & 3.579829 & -30.401569 & 7.040 & 0.22$^{+0.01}_{-0.01}$ & 0.05$^{+0.02}_{-0.02}$ & 0.18$^{+0.01}_{-0.01}$ & 2590$^{+670}_{-640}$ & 34$^{+3}_{-3}$ & 6.62$^{+0.19}_{-0.24}$ & 6.94 \\
J100003.90+020734.2 & 150.016232 & 2.126173 & 7.060 & 2.30$^{+0.12}_{-0.12}$ & 0.08$^{+0.11}_{-0.06}$ & 2.20$^{+0.14}_{-0.14}$ & 2990$^{+380}_{-310}$ & 66$^{+5}_{-4}$ & 7.39$^{+0.11}_{-0.10}$ & 1.00 \\
J021659.48-050907.2 & 34.247828 & -5.151996 & 7.100 & 1.74$^{+0.06}_{-0.06}$ & 1.12$^{+0.13}_{-0.20}$ & 0.64$^{+0.18}_{-0.13}$ & 2150$^{+870}_{-690}$ & 46$^{+13}_{-10}$ & 6.79$^{+0.27}_{-0.30}$ & 1.00 \\
J141943.08+525316.5 & 214.929512 & 52.887913 & 7.155 & 1.19$^{+0.07}_{-0.07}$ & 0.27$^{+0.06}_{-0.06}$ & 0.92$^{+0.08}_{-0.08}$ & 3110$^{+700}_{-740}$ & 57$^{+5}_{-5}$ & 7.21$^{+0.18}_{-0.24}$ & 1.00 \\
J021738.29-051919.4 & 34.409532 & -5.322046 & 7.180 & 2.23$^{+0.25}_{-0.23}$ & 0.37$^{+0.27}_{-0.24}$ & 1.86$^{+0.27}_{-0.29}$ & 2500$^{+920}_{-680}$ & 64$^{+11}_{-11}$ & 7.20$^{+0.27}_{-0.28}$ & 1.00 \\
J021754.78-051712.6 & 34.478255 & -5.286832 & 7.210 & 1.59$^{+0.27}_{-0.26}$ & 0.19$^{+0.12}_{-0.11}$ & 1.39$^{+0.27}_{-0.28}$ & 4520$^{+1280}_{-1220}$ & 419$^{+211}_{-136}$ & 7.64$^{+0.24}_{-0.29}$ & 1.00 \\
J021729.37-050657.3 & 34.372370 & -5.115912 & 7.220 & 1.50$^{+0.13}_{-0.12}$ & 0.28$^{+0.24}_{-0.19}$ & 1.22$^{+0.21}_{-0.26}$ & 1890$^{+600}_{-510}$ & 81$^{+15}_{-18}$ & 6.85$^{+0.23}_{-0.27}$ & 1.00 \\
J033228.58-275333.2 & 53.119075 & -27.892556 & 7.408 & 1.60$^{+0.18}_{-0.15}$ & 0.58$^{+0.19}_{-0.25}$ & 1.05$^{+0.22}_{-0.21}$ & 2350$^{+1050}_{-770}$ & 72$^{+16}_{-15}$ & 7.02$^{+0.31}_{-0.34}$ & 1.00 \\
J095918.44+021420.4 & 149.826836 & 2.239006 & 7.510 & 4.00$^{+0.23}_{-0.21}$ & 0.07$^{+0.10}_{-0.05}$ & 3.92$^{+0.23}_{-0.23}$ & 2780$^{+490}_{-680}$ & 107$^{+8}_{-7}$ & 7.51$^{+0.15}_{-0.24}$ & 1.00 \\
J100138.48+021912.1 & 150.410345 & 2.320026 & 7.620 & 1.30$^{+0.18}_{-0.17}$ & 0.34$^{+0.12}_{-0.14}$ & 0.98$^{+0.16}_{-0.16}$ & 3950$^{+1390}_{-1180}$ & 83$^{+16}_{-15}$ & 7.47$^{+0.28}_{-0.31}$ & 1.00 \\
J095948.35+022655.5 & 149.951441 & 2.448757 & 7.650 & 33.72$^{+0.32}_{-0.31}$ & 5.34$^{+0.31}_{-0.32}$ & 28.38$^{+0.38}_{-0.36}$ & 3860$^{+90}_{-80}$ & 138$^{+2}_{-2}$ & 8.32$^{+0.02}_{-0.02}$ & 1.00 \\
J021735.40-050806.4 & 34.397484 & -5.135116 & 7.760 & 1.30$^{+0.08}_{-0.08}$ & 0.03$^{+0.04}_{-0.02}$ & 1.27$^{+0.08}_{-0.08}$ & 2780$^{+530}_{-680}$ & 45$^{+3}_{-3}$ & 7.24$^{+0.16}_{-0.24}$ & 1.00 \\
J100120.74+021438.8 & 150.336418 & 2.244101 & 8.140 & 4.06$^{+0.31}_{-0.29}$ & 1.30$^{+0.20}_{-0.20}$ & 2.76$^{+0.34}_{-0.34}$ & 3110$^{+480}_{-410}$ & 80$^{+11}_{-11}$ & 7.57$^{+0.13}_{-0.13}$ & 1.00 \\
J095920.97+021329.8 & 149.837357 & 2.224933 & 8.300 & 4.33$^{+0.55}_{-0.53}$ & 1.96$^{+0.27}_{-0.35}$ & 2.42$^{+0.49}_{-0.50}$ & 3790$^{+1630}_{-1410}$ & 114$^{+30}_{-27}$ & 7.71$^{+0.33}_{-0.43}$ & 1.00 \\
J021718.46-051326.2 & 34.326908 & -5.223947 & 8.310 & 4.25$^{+0.29}_{-0.27}$ & 1.38$^{+0.58}_{-0.70}$ & 2.91$^{+0.64}_{-0.59}$ & 1500$^{+550}_{-430}$ & 87$^{+19}_{-18}$ & 6.96$^{+0.25}_{-0.28}$ & 1.00 \\
J141930.27+525251.0 & 214.876139 & 52.880827 & 8.351 & 2.43$^{+0.08}_{-0.07}$ & 1.13$^{+0.10}_{-0.11}$ & 1.30$^{+0.09}_{-0.09}$ & 2260$^{+150}_{-140}$ & 71$^{+5}_{-5}$ & 7.11$^{+0.06}_{-0.06}$ & 1.00 \\
J001433.70-302311.2 & 3.640408 & -30.386438 & 8.500 & 1.52$^{+0.14}_{-0.13}$ & 0.48$^{+0.12}_{-0.11}$ & 1.04$^{+0.17}_{-0.17}$ & 2630$^{+750}_{-670}$ & 150$^{+33}_{-29}$ & 7.20$^{+0.22}_{-0.26}$ & 1.35 \\
J114933.82+222059.2 & 177.390929 & 22.349767 & 8.630 & 3.79$^{+0.31}_{-0.28}$ & 1.66$^{+0.19}_{-0.20}$ & 2.15$^{+0.30}_{-0.30}$ & 3460$^{+1000}_{-900}$ & 115$^{+20}_{-19}$ & 7.63$^{+0.23}_{-0.27}$ & 1.05 \\
J095940.19+021725.5 & 149.917473 & 2.290408 & 8.960 & 2.07$^{+0.34}_{-0.33}$ & 0.17$^{+0.16}_{-0.11}$ & 1.89$^{+0.36}_{-0.36}$ & 2790$^{+890}_{-760}$ & 99$^{+24}_{-22}$ & 7.43$^{+0.26}_{-0.30}$ & 1.00 \\
J100032.70+021828.9 & 150.136253 & 2.308030 & 9.288 & 1.50$^{+0.14}_{-0.13}$ & 0.04$^{+0.05}_{-0.03}$ & 1.45$^{+0.14}_{-0.14}$ & 2940$^{+660}_{-710}$ & 70$^{+8}_{-7}$ & 7.43$^{+0.19}_{-0.24}$ & 1.00 \\
\enddata
\tablecomments{All given error bounds reflect only the uncertainties derived from the Bayesian line fitting, and do not include additional systematic uncertainties. A subscript $n$ indicates a narrow line component and a subscript $b$ indicates a broad-line component. All values are corrected for magnification from gravitational lensing (when appropriate). We detail the sources of the spectra and previous references for each object in Table~\ref{tab:apdx}.}
\end{deluxetable*}

\subsection{Black Hole Masses}\label{sec:bhms}
Measurements of black hole masses at high redshifts ($z\gtrsim2$) are currently uncertain. While studies leveraging reverberation mapping at $z\lesssim1$ have produced scaling relations that relate the luminosity and kinematic width of broad-lines observed in AGN broad-line regions to the mass of the black hole \citep[e.g.][]{greene05,reines13,reines15,dallabonta25}, the accuracy and applicability of these relations at high-redshift is hotly debated. Even for non-LRD objects that resemble local BLAGN, differences in Eddington ratios \citep[e.g.][]{Du2016, FonsecaAlvarez2020}, broad-line region sizes, and black hole environments may produce deviations of $\sim 0.7$~dex from the virial relations \citep{abuter24}. 

For LRDs, these uncertainties are even greater, as multiple models suggest that the broad-lines observed in LRDs do not probe the broad-line region directly, and instead are the result of scattering within a dense medium \citep[e.g.][]{rusakov25}. \cite{greene26} have also proposed that standard bolometric corrections may be inappropriate for LRDs, casting further doubt upon the accuracy and applicability of virial relations---which also depend on line luminosity---to LRDs. However, a recent study \citep{juodzbalis26} has shown that for one lensed, multiply imaged LRD, the virial relations of \cite{reines15} are well matched to a black hole mass derived from spectroastrometry. Given the unsettled nature of this debate, we fall back to the traditional approach and use the virial relation presented in Equation~10 from \cite{greene05}, reproduced here as Equation~\ref{eq:bhmass}, where $L_{{\rm H}\beta}$ and $\textrm{FWHM}_{{\rm H}\beta}$ are the luminosity and FWHM of the broad component of \Hb{}, respectively. 
\begin{equation} \label{eq:bhmass}
\begin{aligned} 
M_{\textrm{BH}}&=(2.4\pm0.3) \\ & \times 10^6\left(\frac{L_{{\rm H}\beta}}{10^{42}\textrm{ erg s}^{-1}}\right)^{0.59\pm0.06} \\ & \times \left(\frac{\textrm{FWHM}_{{\rm H}\beta}}{10^3\textrm{ km s}^{-1}}\right)^{2}M_{\odot} 
\end{aligned}
\end{equation}

We note that this equation assumes that the \Hb{} line has been corrected for extinction. As our sources (by design) lie at $z>7$, \Ha{} is not available for such a correction, and the PRISM resolution is too poor to robustly de-blend the H$\gamma$ line from the [O\,\textsc{iii}]$\lambda$4363 line. Therefore we cannot perform an extinction correction. Moreover, even if \Ha{} or H$\gamma$ were available, LRDs and BLAGN at $z>3$ have shown strong differential reddening of the broad and narrow components of their Balmer lines \citep[e.g.][]{brooks25}, and the dense gas scenarios proposed by \cite{naidu25} and \cite{inayoshi25} naturally create Balmer line ratios inconsistent with standard Case B recombination assumptions, even in the absence of dust. As such, we do not attempt an extinction correction for \Hb{}. If \Hb{} is indeed attenuated, our black hole masses may be underestimated. However, because $M_{\textrm{BH}}$ only scales with $L_{\textrm{H}\beta}^{0.59}$, this uncertainty is likely small compared to the systematic $\sim0.7$~dex uncertainty in $M_{\textrm{BH}}$ \citep{abuter24}.

Our sample spans a range of black hole masses from $4\times10^{6}-2\times10^8$~$M_{\odot}$ with the lower bound primarily set by the minimum FWHM detectable with the PRISM (see \S\ref{sec:completeness} below). 

\subsection{LRD Selection} \label{sec:lrdselect}

As summarized in \S\ref{sec:intro}, the exact physical nature of LRDs, and even the specific defining observed characteristics of LRDs are hotly debated. Despite these debates, several general criteria for selecting LRDs from photometric or spectroscopic data have emerged. In more data-rich spectroscopy, rest-frame UV and rest-frame optical continuum slopes (\betauv{} and \betaopt, respectively) are frequently the basis of common selection criteria, with most studies requiring \betauv$\lesssim0$ and \betaopt$\gtrsim0$, thereby producing the typical LRD ``v-shaped'' SED \citep[e.g.][]{kocevski24,akins25c,barro24b}. 
The other key selection criterion derived from photometry is source compactness in the rest-frame optical, typically probed using F444W imaging.

For this study, we define an LRD as an object with \betauv$<0$ (measured from the PRISM spectra with strong emission lines masked at rest-frame $1500-3645$\AA{}), \betaopt$>0$ (measured from the PRISM spectra with strong emission lines masked at rest-frame $4000-6500$\AA{}), and a F444W half-light radius $<$0\farcs09 derived from coordinate-matched photometry (roughly corresponding to the stellar size-magnitude locus, e.g. \citealt{kocevski25}). These criteria are deliberately chosen as a compromise between \cite{kocevski25}, who require \betauv$<-0.37$ and \betaopt$>0$ and \cite{akins25c}, who instead require only a red color cut $m_{\textrm{F277W}}-m_{\textrm{F444W}}>1.5$, where $m_{\textrm{F277W}}$ and $m_{\textrm{F444W}}$ are F277W and F444W apparent magnitudes. 18/20 objects in this study satisfy these criteria. Two objects---J114933.82+222059.2 and J021718.46-051326.2---do not pass this criterion, and exhibit \betaopt$<0$. However, J114933.82+222059.2 was first introduced as an LRD in \cite{tripodi25}, but just barely fails our \betaopt$>0$ criterion, exhibiting \betaopt$=-0.22^{+0.34}_{-0.32}$. As a result we also classify this object as an LRD as its 1$\sigma$ upper limit on \betaopt{} meets our criteria. J021718.46-051326.2 is a more complicated case, as it is in fact a superposition of a $z=8.31$ LRD with a blue, lower-redshift interloper, identified both in imaging and via off-center flux blue-ward of the Ly$\alpha$ break of the LRD. As a result, despite only selecting for broad-\Hb{} emission and not requiring any LRD-specific cuts, our 20 object sample consists entirely of LRDs. 

\subsection{Sample Completeness}\label{sec:completeness}

To prepare for the construction of the black hole mass function, we consider two main sources of incompleteness in our sample: (1) our ability to detect and select a broad-\Hb{} line in spectroscopic data (referred to hereafter as ``line detection completeness'') and (2) the fraction of photometrically identified objects in our surveyed fields that have spectroscopy (referred to hereafter as ``observed fraction completeness'').

\subsubsection{Line Detection Completeness}\label{sec:detcomp}
 
To address detection completeness, we iterate on the methodology presented in \cite{taylor25a}. We simulate \Hb{} and [O\,\textsc{iii}]$\lambda\lambda4959,5007$ lines as a function of the broad-\Hb{} FWHM, broad-\Hb{} line flux, narrow-\Hb{} line flux, and redshift (and thus when combined: black hole mass). We use the \texttt{pandeia} engine \citep{pontoppidan16}---the engine behind the \jwst{} Exposure Time Calculator---to simulate NIRSpec/PRISM observations of these objects on a grid of exposure times that span the exposure times in the dataset. We then use the same line fitting algorithms described in \S\ref{sec:fitting} to attempt to fit and re-select the simulated objects. 

\begin{figure}[t]
\includegraphics[width=\columnwidth]{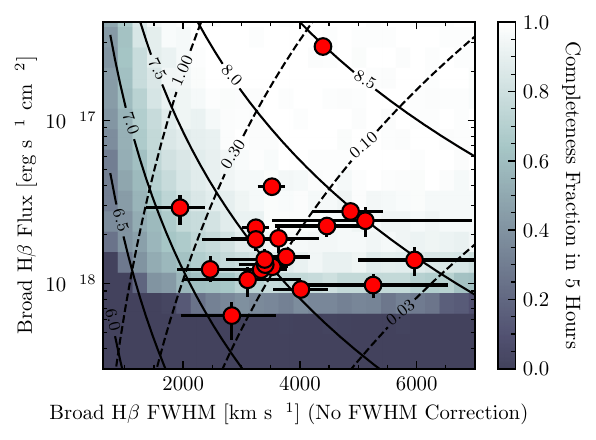}
\includegraphics[width=\columnwidth]{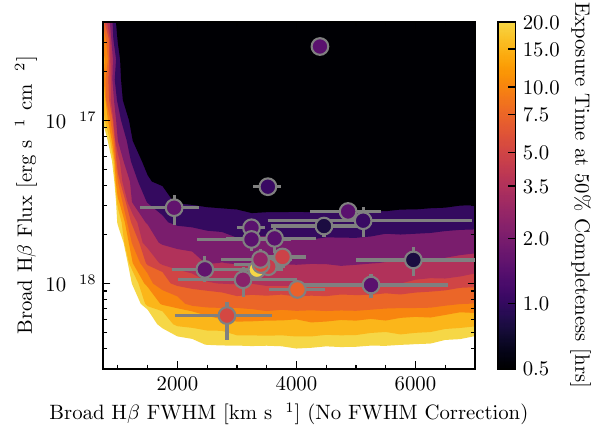}
\caption{\textit{Upper Panel}: Completeness (colored shading) for a five hour PRISM exposure as a function of broad-\Hb{} flux and broad-\Hb{} FWHM at $z=8.5$. Solid overplotted lines are contours of logarithmic black hole mass (in solar masses) \citep[as computed with][]{greene05}, and dashed overplotted lines are contours of Eddington ratio (assuming unattenuated Case B recombination [\Ha{}/\Hb{}$=2.86$] and a 130$\times$ bolometric correction from \Ha{} flux to bolometric luminosity). Our sample is highly detection complete above $10^{7.5}$\,$M_{\odot}$, and is consistent with Eddington ratios $\leq1$. \textit{Lower panel}: Contours of exposure time corresponding to 50\% completeness in broad-\Hb{} flux and broad-\Hb{} FWHM space at $z=8.5$. Here, both contours and colored points (the BLAGN sample) are colored by exposure time. Points that are a darker shade than their overlapping contours are objects with significantly lower completeness, while objects with lighter shades than overlapping contours are highly complete. As expected, increased exposure times offer sensitivity to fainter line fluxes, and have little to no effect on FWHM sensitivity.} \label{fig:detcompleteness}
\end{figure}

Our completeness simulations reaffirm that our ability to detect BLAGN with NIRSpec/PRISM is limited by both line flux and the spectral resolution of the PRISM. In the upper panel of Figure~\ref{fig:detcompleteness}, we plot completeness for a five hour PRISM exposure as a function of broad-\Hb{} flux and broad-\Hb{} FWHM at $z=8.5$. Figure~\ref{fig:detcompleteness} demonstrates that---for a five hour PRISM observation---BLAGN with broad-\Hb{} fluxes $<1\times10^{-18}$\,erg\,s\per\,cm\persq{} are almost entirely undetected at all broad-\Hb{} FWHMs. However, at brighter line fluxes, the dominant limit for completeness is instead the broad-\Hb{} FWHM. We overplot contours of $\log\left(M_{\textrm{BH}}/M_{\odot}\right)$ as solid curves in the upper panel of Figure~\ref{fig:detcompleteness}. Conveniently, these contours well trace the shape of the completeness shading at low FWHM. As such, for a five hour PRISM exposure, the completeness reaches $50\%$ near $M_{\textrm{BH}}\approx 10^7\,M_{\odot}$, down to the line flux limit. 

Within the \cite{greene05} formalism, moving along a $M_{\textrm{BH}}$ contour in Flux/FWHM space necessitates a change in Eddington ratio $(f_{\textrm{Edd}})$. We overplot contours of Eddington ratio as dashed curves in the upper panel of Figure~\ref{fig:detcompleteness}. As $M_{\textrm{BH}}$ scales with both FWHM and line flux, at fixed $M_{\textrm{BH}}$, a narrower FWHM will necessitate a brighter line, and thus a higher Eddington ratio. We note that all of our BLAGN (even without the LRD-specific bolometric correction prescribed in \citealt{greene26}) exhibit $f_{\textrm{Edd}}<1$ (assuming accurate masses from the single-epoch scaling relations). The $f_{\textrm{Edd}}<1$ contours also help explain why we fail to identify any sources at broad-\Hb{} fluxes $\gtrsim 4\times10^{-18}$\,erg\,s\per\,cm\persq{} and pre-correction FWHM\,$<$\,2000\,km\,s\per{}. This lack of sources is not in fact due to the PRISM resolution; rather, such sources would exhibit super-Eddington accretion. We also use the estimated Eddington ratios to bound the completeness simulations. Naively, when considering---for example---the completeness for BLAGN with $M_{\textrm{BH}}=10^7\,M_{\odot}$, our data are entirely insensitive to a hypothetical object with broad-\Hb{} line flux $\sim1\times10^{-20}$\,erg\,s\per\,cm\persq{} and with broad-\Hb{} FWHM $\sim8300$\,km\,s\per{}, and thus our completeness correction should compensate for this insensitivity. However, this hypothetical object would also have an extremely low $f_{\textrm{Edd}}\sim0.0026$. To avoid over-correcting for such likely nonphysical objects, we bound our BLAGN sample and the completeness corrections between $f_{\textrm{Edd}}=0.03-1.00$, which more than span our current sample.  

We plot the dependence of our line detection completeness on exposure time in the lower panel of Figure~\ref{fig:detcompleteness}. Here, we plot contours corresponding to $50\%$ completeness across the same broad-\Hb{} flux and broad-\Hb{} FWHM axes as the upper panel. As expected, increased exposure times offer greater sensitivity to fainter line fluxes. Some objects in our sample (with darker shaded points than the overlapping background contours) are unexpected finds: objects detected with lower exposure times than otherwise expected for their broad-\Hb{} flux and broad-\Hb{} FWHM. These objects occupy (for their individual exposure times) regions of low completeness, and thus large corrective factors are applied to account for this. Inversely, objects shaded much lighter than the overlapping contours are in regions of very high completeness, could have been detected with much less exposure time, and have only small completeness correction factors. 

When considering lensed objects, we base the detection completeness on their masses computed from their observed-frame line flux and FWHM, as this best reflects the recovery rate of \Hb{} broad-lines in spectra with the same observed line properties as the observed lensed objects. 

\begin{figure}[t]
\includegraphics[width=\columnwidth]{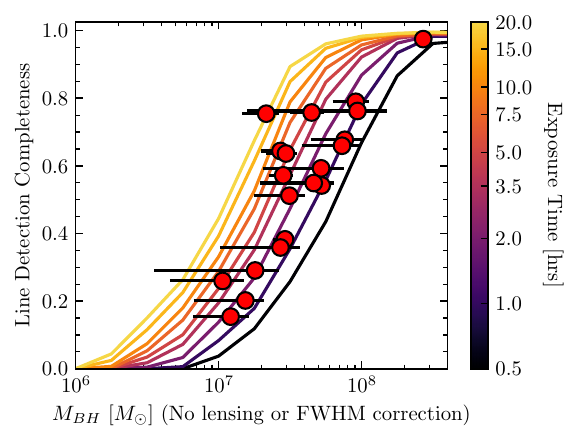}
\caption{Line Detection completeness fraction as a function of black hole mass for different exposure times. As expected, longer exposure times result in higher completeness due to increased sensitivity to lower broad-line fluxes. The majority of the sample is selected from relatively shallow ($\lesssim2$\,hours) data, suggesting that much higher completeness and larger samples of BLAGN may be found in future deeper observations.}
\label{fig:detcompleteness_lines}
\end{figure}

We derive our final per-exposure-time completeness curves by averaging the completeness in broad-\Hb{} flux and broad-\Hb{} FWHM space along $M_{\textrm{BH}}$ contours at $0.03<f_{\textrm{Edd}}<1.00$. We show these curves in Figure~\ref{fig:detcompleteness_lines}.

\subsubsection{Observed Fraction Completeness}\label{sec:selcomp}

While \jwst{}/NIRSpec/PRISM spectroscopic surveys have made incredible strides in recent years, achieving high observed fractions ($f_{\textrm{obs}}$) of photometric sources remains incredibly difficult, even in well studied fields. While our line detection completeness compensates for objects observed in shallow data with insufficient depth to detect broad-lines, in computing the black hole mass function we must also correct for photometrically identified objects that do not yet have PRISM spectra---regardless of exposure time. When compared to the line detection completeness simulations, this correction is straightforward, with a few important caveats. 

While simply computing the fraction of photometric objects with spectra as a function of brightness (e.g. F444W magnitude) would be a straightforward correction, it fails to account for selection biases in the design of NIRSpec programs. For example, both RUBIES and EMBER deliberately targeted bright and red objects, with EMBER placing a specific emphasis on objects with photometric signatures consistent with LRDs. While it is impractical to characterize the selection functions (or lack thereof) of all of the programs in our sample, it is insufficient to ignore selection effects entirely. 

\begin{figure}[t]
\includegraphics[width=\columnwidth]{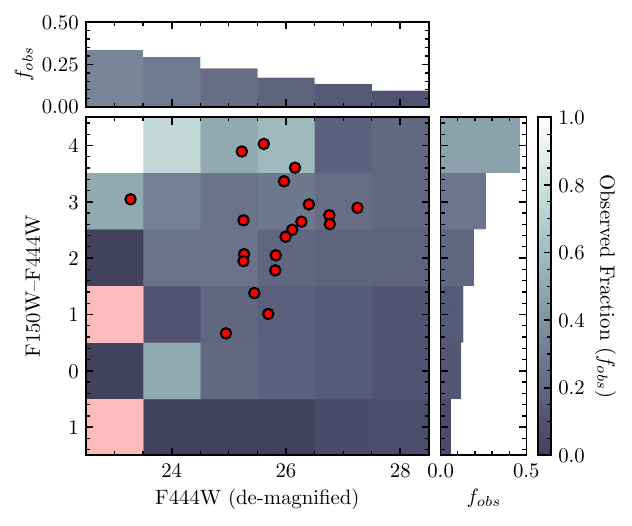}
\caption{Observed fraction ($f_{\textrm{obs}}$) completeness as a function of F444W magnitude and F150W--F444W color. Areas colored pink do not contain any photometric or spectroscopically observed objects. We plot $f_{\textrm{obs}}$ as a function of F444W magnitude and F150W--F444W color in the bar plots on the top and right axes. As expected from the inclusion of surveys that targeted red objects (e.g. RUBIES, EMBER), $f_{\textrm{obs}}$ rises toward redder colors. Computing $f_{\textrm{obs}}$ in color--magnitude space compensates for this bias. These observed fraction completeness measures are ultimately multiplied by the line detection completeness measures to give final overall completeness correction factors for each source.}
\label{fig:obscomplete}
\end{figure}

To mitigate these biases, we first down-select from the $2,001,075$ photometric objects identified from the UNICORN project across all of the surveyed fields with F150W and F444W data by considering only objects with best-fit photometric redshifts (or spectroscopic redshifts) $7.0<z<9.5$, matching the spectroscopic sample. As these selection effects are typically dominated by brightness and color, we compute $f_{\textrm{obs}}$ as a function of both F444W magnitude and F150W--F444W color. We show $f_{\textrm{obs}}$ in this color--magnitude space in Figure~\ref{fig:obscomplete}. To account for the effects of lensing, we compute $f_{\textrm{obs}}$ using de-magnified F444W magnitudes. This allows us to properly compute $f_{\textrm{obs}}$ based on the intrinsic brightnesses of the objects, independent of lensing. The observed fractions of the sample range from 0.16--0.57. 

\section{Results} \label{sec:results}

\subsection{The Black Hole Mass Function}\label{sec:BHMF}

\begin{figure}[t]
\includegraphics[width=\linewidth]{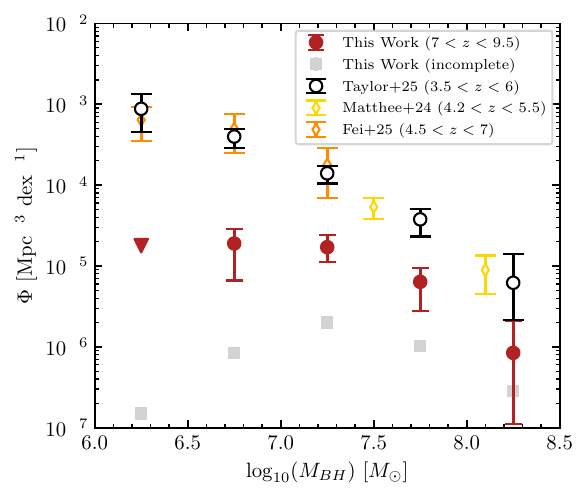}
\caption{\jwst{}-based \bhmf{}s at $z\sim5$ \citep[hollow points;][]{taylor25a,matthee24,fei25} and $z\sim8$ (solid points). For this work, we show both the completeness corrected points as solid red circles, and the equivalent \bhmf{} without a completeness correction as grey squares. The \bhmf{} shows significant ($>3.5\sigma$) evolution over $z\sim5-8$ in all but the two heaviest mass bins where number statistics limit the significance to $2.0\sigma$ and $1.4\sigma$ respectively.}
\label{fig:bhmf}
\end{figure}

Here we use the previously introduced BLAGN sample and completeness corrections to compute the black hole mass function. After correcting for the effects of gravitational lensing (when applicable), we determine a total combined photometric surveyed area of 1942~arcmin$^{2}$ (0.54~degrees$^{2}$), corresponding to a comoving volume of $9.30\times10^6$~cMpc$^3$ when integrated over $7.0<z<9.5$. Here, our completeness corrections allow us to assume a flat selection function in redshift space. 

To best consider the uncertainties in \mbh{} derived from uncertainties in the broad-\Hb{} line measurement (excluding systematic uncertainties), we treat each realization of \mbh{} from the posteriors measured during the MCMC fitting as an individual BLAGN. We then weight each posterior sample by $1/N_{\textrm{samples}}$. In effect, each BLAGN is fractionally distributed into the mass bins of the \bhmf{} weighted by its \mbh{} posterior. Here each \mbh{} posterior measurement is independently corrected for line detection and observed fraction completeness. We show our final $7.0<z<9.5$ \bhmf{} as the solid red points in Figure~\ref{fig:bhmf} and provide the data in Table~\ref{tab:bhmfz7all}. 

As a standard of comparison, we also re-compute the \bhmf{} derived in \cite{taylor25a}. Here, while we keep the same sample, redshifts, and black hole masses, we take advantage of our improved observed fraction completeness scheme, updated photometry, and improved photometric redshifts. For brevity, we label these points as ``Taylor+25'' in subsequent figures. We provide the data behind this reprocessed mass function in Appendix~\ref{apdx:table}.

Comparing the normalizations of the full (LRD+non-LRD) \bhmf{}s, we find a $\approx1.0$~dex increase in the \bhmf{} from $z\sim8-5$, consistent with moderate black hole growth over the $\sim625$~Myr period between the two epochs. Alternatively, the two mass functions can be converged by shifting the $z\sim8$ mass function by $\approx0.6$~dex toward heavier masses until it coincides with the $z\sim5$ mass function. In the simplest of models, this $\approx0.6$~dex increase can be interpreted as steady black hole growth over $\sim625$~Myr at a constant $f_{\textrm{Edd}}=0.1$. While this model is over-simplified, it nonetheless affirms that the evolution of the \bhmf{} is easily achievable in the time between the measured epochs with a standard sub-Eddington growth paradigm. 

\begin{deluxetable}{cccccc}[h]
\tablecaption{$7<z<9.5$ Black Hole Mass Function (all LRDs)} \label{tab:bhmfz7all}
\tablehead{$\log\left(M_{BH}/M_{\odot}\right)$ & $\Phi$ [$\times10^{-6}$] & N & \multicolumn{3}{c}{Completeness} \\
 & Mpc$^{-3}$\,dex$^{-1}$ & & (total) & (obs) & (det)}
\startdata
6.25 & $<17.7$ & 0.68 & 0.02 & 0.22 & 0.07 \\
6.75 & $19.1^{+9.6}_{-12.5}$ & 3.93 & 0.04 & 0.23 & 0.19 \\
7.25 & $17.1^{+7.1}_{-5.9}$ & 9.27 & 0.12 & 0.26 & 0.45 \\
7.75 & $6.36^{+3.19}_{-3.58}$ & 4.77 & 0.16 & 0.24 & 0.66 \\
8.25 & $0.84^{+1.25}_{-0.73}$ & 1.33 & 0.34 & 0.36 & 0.95 \\
\enddata
\tablecomments{``det'' refers to line detection completeness, ``obs'' refers to observed fraction completeness, and ``total'' is the product of  ``det'' and ``obs''}
\end{deluxetable}

\subsection{The Bolometric Luminosity Function and Black Hole Accretion Density}\label{sec:BHAD}

\begin{figure*}[t]
\includegraphics[width=0.49\textwidth]{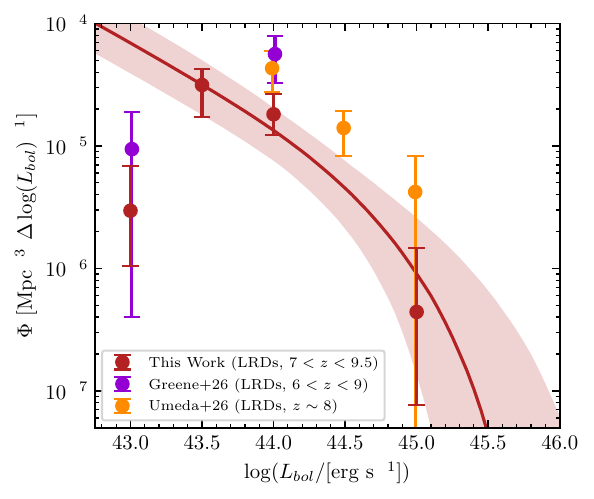}
\includegraphics[width=0.49\textwidth]{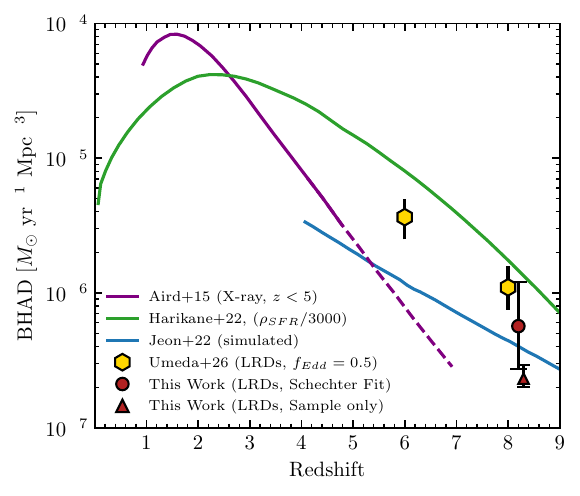}
\caption{Left panel: The $L_{\textrm{bol}}$ luminosity function for the full $z\sim8$ \Hb{}-emitter sample computed using the \cite{greene26} $L_{\textrm{bol}}/L_{5100}$ correction, as well as recent literature results from LRDs. This luminosity function falls off rapidly at $L_{\textrm{bol}}\sim10^{45}$~erg~s\per{} and appears to extend in a power-law-like manner to lower luminosities (omitting the lowest luminosity highly incomplete data point). The red curve and shaded region shows the median and 68\% confidence interval Schechter function fit to the most luminous three data points from this work. Right panel: The BHAD as a function of redshift. We also overplot the cosmic star-formation rate density from \cite{harikane22} re-normalized by a factor of 3000 to match the BHAD at $z\sim3$ as a point of reference. Our detected BLAGN sample produces a result that broadly agrees with a photometric LRD sample from \cite{umeda26}. This sample was initially introduced in \cite{akins25c} using non-LRD specific assumptions and was reprocessed with LRD-specific corrections by \cite{umeda26}. We find good agreement with the BHAD derived from the $M_{H,L}=10^8 M_{\odot}$ models of \cite{jeon22}, a simulation in which BHs are seeded and grow within atomic cooling halos. We note that our data are likely a lower limit on the true BHAD, as they do not account for obscured, non-broad-line AGN.}
\label{fig:Lbollf}
\end{figure*}

We next compute and analyze the bolometric luminosities of the broad-\Hb{} emitter sample. We first measure $L_{5100}$ directly from the flux-calibrated PRISM spectra by averaging the spectral flux over 5050--5150 rest-frame \AA{}. We then estimate the bolometric luminosities $(L_{\textrm{bol}})$ for the sample using the correction $L_{\textrm{bol}}/L_{5100}=5.4$ from \cite{greene26}. 
We use these \lbol{} measurements to construct a bolometric luminosity function using the same comoving volume and per-source completeness as the \bhmf{}. We show the resulting \lbol{} luminosity function in Figure~\ref{fig:Lbollf}. 

The \lbol{} luminosity function shows a power-law-like faint end, and a steepening decline at \lbol{}$>10^{44.5}$~erg~s\per{}. Given this visual form, we next fit a Schechter function \citep{schechter76} using MCMC to the four most luminous (and complete) data points in the luminosity function, and show the median and 68\% confidence interval for the resulting fit as the red curve in Figure~\ref{fig:Lbollf}. Due to the small number of points in this fit, to ensure fit stability, we fix the faint end slope to $\alpha=-0.65$ following from the $z=4-5$ X-ray data of \cite{ueda14}, and require that $L^*<10^{46}$~erg~s$^{-1}$, motivated by the bright-end luminosity limit of our data. 

Using this Schechter fit, we integrate over the luminosity function from $42<\log{(L_{\textrm{bol}}/[\textrm{erg s}^{-1}])}<\infty$ and---assuming a radiative efficiency of $\epsilon=0.1$, as is typical in the literature \citep[e.g.,][]{umeda26}---compute the Black Hole Accretion Density (BHAD) at $z\sim8$. We find that the BHAD~$=4.8^{+5.8}_{-2.6}\times10^{-7}~M_{\odot}~\textrm{yr}^{-1}~\textrm{Mpc}^{-3}$. As the faint-end slope is shallow and fixed at $\alpha=-0.65$, the uncertainty on this value is driven by the large error bars for the most luminous points (due to low number statistics) which limit the constraints on the Schechter fit at the bright end of the luminosity function. If we instead simply sum the completeness-corrected luminosities of our sample (deriving errors via bootstrapping), we compute BHAD~$=2.3^{+0.6}_{-0.3}\times10^{-7}~M_{\odot}~\textrm{yr}^{-1}~\textrm{Mpc}^{-3}$. It is important to note that these results are lower bounds on the full BHAD in this epoch, as this study only includes broad-line emitting objects and is insensitive to obscured AGN and fainter BLAGN. These results---the first based on a spectroscopic sample at $z\sim8$---show broad agreement with the photometric LRD data from \cite{umeda26}. \cite{umeda26} use the same base sample as \cite{akins25c}, but use updated assumptions about LRD bolometric luminosities, dust content, and assume $f_{\textrm{Edd}}=0.5$. Our data also offer possible agreement with the simulations of \cite{jeon22}, but disagree with an extrapolation of pre-\jwst{} X-ray detected AGN sources at $z<5$ \citep{aird15}. We show these results in the right panel of Figure~\ref{fig:Lbollf}. Our BHAD excess over the X-ray extrapolation further reinforces the growing evidence that SMBH growth in the early universe was much faster than previously expected, producing massive SMBHs quickly within only the first $\sim750$~Myr of cosmic time. 

As a byproduct of constructing the bolometric luminosity function and BHAD, we also compute Eddington ratios for the sample by comparing their bolometric luminosity-derived accretion rates to Eddington rates computed from the sample's \mbh{}s (assuming that the LRD masses are consistent with virial relations). We find that the completeness-corrected mean Eddington ratio of the sample is $f_{\textrm{Edd}}=0.08$. This quantity is very comparable to the aforementioned rough estimate of $f_{\textrm{Edd}}=0.1$ (see \S\ref{sec:BHMF}) needed to reproduce the BLAGN \bhmf{} at $z\sim5$ starting from the \bhmf{} at $z\sim8$ with constant, fixed-$f_{\textrm{Edd}}$ black hole growth. While such steady growth is unlikely, and more bursty models of accretion are favored in recent models \citep[e.g.][]{jeon25}, this mean marginalizes over this burstiness, further indicating that while super-Eddington accretion (or heavy black hole seeds) may be necessary to produce the observed \bhmf{} at $z\sim8$ (see \S\ref{sec:models}), evolution from $z\sim8-5$ can be achieved with standard accretion mechanisms. 

\subsection{The UV Luminosity Function}\label{sec:LFs}

\begin{figure}[t]
\includegraphics[width=\columnwidth]{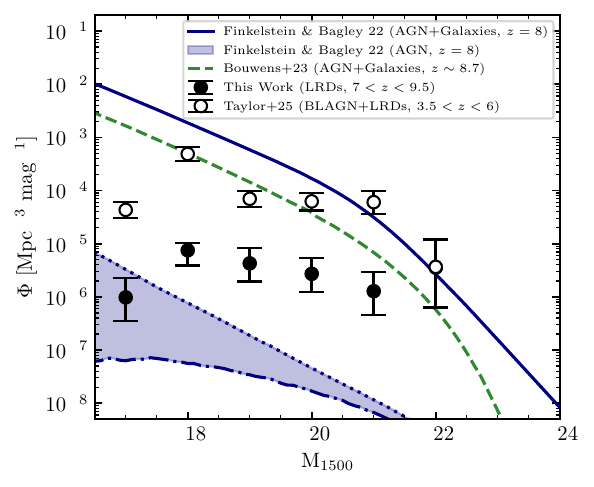}
\caption{The UV luminosity functions of the $z\sim8$ BLAGN sample in this work and the $z\sim5$ BLAGN sample from \cite{taylor25a} compared to pre-\jwst{} luminosity functions from \cite{finkelsteinbagley22}. We plot the best fit galaxy+AGN UV luminosity function at $z=8$ from \cite{finkelsteinbagley22} as a solid dark blue curve, and we shade the region between their AGN best-fit curve where the AGN UV luminosity function was permitted to turn over (dashed dark blue curve) and their AGN best-fit curve where the AGN UV luminosity function was fit as a power-law to faint magnitudes (dotted dark blue curve). We plot the early-\jwst{} galaxy+AGN UV luminosity function from \cite{bouwens23} at $z=8.7$ as the dashed green curve.
At both $z\sim5$ and $z\sim8$, BLAGN are $\sim1-2$~dex more abundant at all UV luminosities than pre-\jwst{} fits predicted, and the $z\sim8$ objects appear to make up $\sim2-10\%$ of the objects at the bright end of the luminosity function.}
\label{fig:uvlf}
\end{figure}

We next measure the absolute UV magnitudes of the BLAGN sample and construct the UV luminosity function. We measure UV magnitudes directly from the flux-calibrated PRISM spectra by averaging the spectral flux over 1450--1550 rest-frame angstroms and applying the appropriate distance modulus to each source. We then construct the UV luminosity function, again using the same comoving volume and per-source completeness as the \bhmf{}. We plot the resulting UV luminosity function, and the UV luminosity function from \cite{taylor25a} (reprocessed with our updated completeness method described in \S\ref{sec:completeness}) in Figure~\ref{fig:uvlf}. 

We compare the observed BLAGN UV luminosity functions to the pre-\jwst{} luminosity functions from \cite{finkelsteinbagley22} and the early-\jwst{} luminosity function from \cite{bouwens23}, and find that at both $z\sim5$ and $z\sim8$, the observed BLAGN luminosity functions exceed the pre-\jwst{} AGN UV luminosity functions by $\sim1-2$~dex (Figure~\ref{fig:uvlf}). This was previously reported at $z\sim5$ by \cite{harikane23,maiolino24,kokorev24,kocevski25,taylor25a} and we now confirm this spectroscopically at $z\sim8$. Moreover, we find that at the bright end, our LRDs make up $\sim2\%$ of the brightest objects when compared to \cite{finkelsteinbagley22}, and $\sim10\%$ of the normalization of \cite{bouwens23}. 

\section{Discussion} \label{sec:discussion}

\subsection{LRDs and the Black Hole Mass Function}\label{sec:bhmf_lrds}
\begin{figure*}[t]
\includegraphics[width=0.49\textwidth]{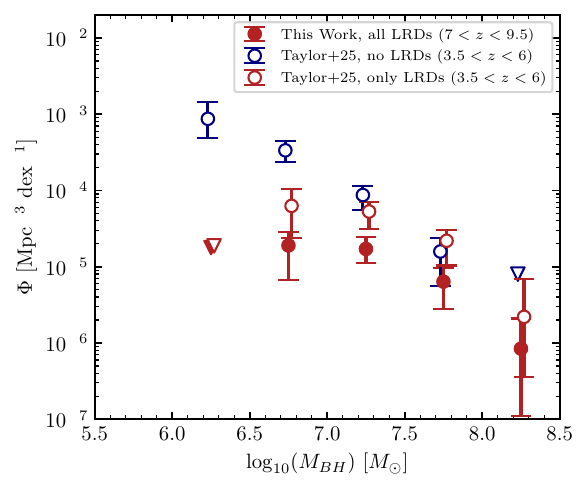}
\includegraphics[width=0.49\textwidth]{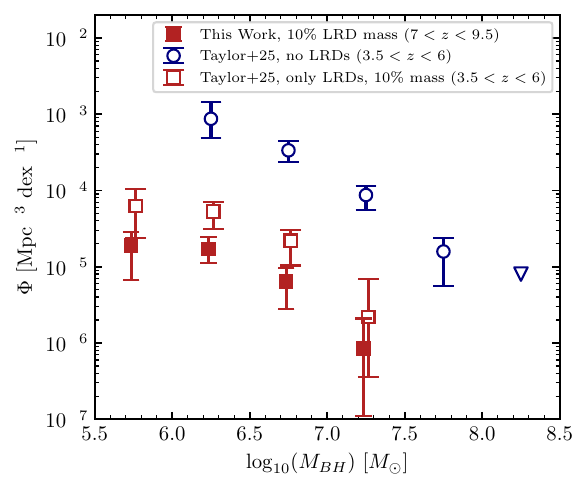}
\caption{Left: The reprocessed \bhmf{}s from \cite{taylor25a} and this work separated into LRDs and non-LRDs. Downward pointing triangles indicate $1\sigma$ upper limits in mass bins occupied by fewer than one object (see \S\ref{sec:BHMF}). The non-LRDs are not present at $z\sim8$, and thus show significant evolution, while the LRDs show only moderate evolution. Right: Here, we decrease all of the LRD masses by a factor of 10 in an approximation of the results of \cite{rusakov25}. At $z\sim5$, this modification causes the \bhmf{} to be completely dominated by non-LRDs at all masses.}
\label{fig:bhmf_lrd} \label{fig:bhmf_lrd_0.1}
\end{figure*}

As previously introduced in \S\ref{sec:intro}, the presence of LRDs---and the associated uncertainties in their black hole masses and status as BLAGN---significantly complicates the interpretation of the \bhmf{}. The applicability of AGN-derived scaling relations for estimating LRD masses is hotly debated \citep[e.g.][]{rusakov25,juodzbalis26}, and the physical interpretation of LRDs as AGN remains uncertain. While we have previously assumed in this work that the broad \Hb{} lines emitted by LRDs directly trace AGN broad-line region kinematics, we now explore alternatives to this assumption. 

The most extreme alternative case is to simply reject the idea of LRDs being or hosting AGN at all. In this case, the broad-\Hb{} lines would have no predictive power on black hole masses whatsoever, and LRDs would therefore not belong in any \bhmf{}. This case is explored in Figure~\ref{fig:bhmf_lrd} (left panel), where we have split the \bhmf{}s of this work and \cite{taylor25a} into separate mass functions for LRDs and non-LRDs. If LRDs are not AGN at all, the $z\sim5$ \cite{taylor25a} \bhmf{} should be replaced by the blue (non-LRD) points from Figure~\ref{fig:bhmf_lrd} (left panel) and the $z\sim8$ \bhmf{} would become purely volume- and completeness-limited upper limits. While the \cite{taylor25a} sample of broad-\Ha{} emitters at $3.5<z<6.0$ consisted of $37\%$ LRDs by number counts, our $7.0<z<9.5$ sample is entirely made up of LRDs. This trend was also observed in \cite{kocevski26}, where the broad-line LRD fraction reaches 100\% at $z=7$. Such a scenario severely limits our ability to comment on the evolution of the number density of non-LRDs at $z >$ 7 as there are simply no detections in our sample. Moreover, it would require extremely rapid evolution of the abundance of observable growing SMBHs between $z\sim8$ and $z\sim5$.

Let us then consider a different treatment for the LRDs. Instead of excluding them entirely, let us assume that LRDs do indeed host AGN, but their observed broad-lines do not reliably trace BLR kinematics, resulting in black hole masses that are overestimated by a factor of ten \citep[e.g.][]{rusakov25}. This rescaling, when averaged across the completeness weighted sample, is equivalent to computing \mbh{} for the sample using bolometric luminosities (using the formalism of \citealt{greene26}, see \S\ref{sec:BHAD}), and assuming a 10\% radiative efficiency and $f_{\textrm{Edd}}=0.8$. For reference, assuming $f_{\textrm{Edd}}=0.5$, as in \cite{umeda26}, would instead reduce our completeness weighted \mbh{} to 16\% of our fiducial masses, and shift our \bhmf{} lighter (to the left) by 0.8~dex. 

We explore the factor of ten reduction in \mbh{} in Figure~\ref{fig:bhmf_lrd_0.1} (right panel). Here, at $z\sim5$, the LRDs contribute less than 10\% of the number density of BLAGN at all masses, making their contributions nearly negligible. As a result, the $z \sim$ 5 \bhmf{}'s normalization falls by $\sim0.3$ dex at the heavy end, and is nearly unchanged at the light end. Due to the 100\% fraction of LRDs at $z\sim8$, applying a factor-of-ten mass adjustment drastically alters the \bhmf{}, reducing its normalization by 0.5~dex. 

Returning to our fiducial assumption that the broad \Hb{} lines emitted by LRDs directly trace AGN broad-line region disk kinematics, separating the \bhmf{} into its LRD and non-LRD components also offers insight into LRDs as a population. Here, even if the \mbh{} measurements for LRDs are inaccurate, we are comparing LRDs to LRDs using the same methodology, thus any global inaccuracies in \mbh{} should in principle cancel out in the comparison (provided that the scalings between observable and physical properties remain the same for LRDs at all epochs). The first line of intrigue is the very strongly evolving fraction of broad-line emitters at \mbh{}$>10^7\,M_{\odot}$ that are LRDs ($52\%$ at $3.5<z<6.0$ versus $100\%$ at $7.0<z<9.5$). As our sample objects (and those of \citealt{taylor25a}) are simply identified by their broad-lines, with no selection parameters that prefer LRDs, these percentages are not biased by our broad-line detection techniques. However, the spectroscopically observed fraction of each type of source does play a role in determining these percentages. If we correct these percentages using our observed fraction completeness measures, we find that LRDs make up $48\%$ of broad-line emitters at $3.5<z<6.0$ and---in this sample---$100\%$ at $7.0<z<9.5$, showing good agreement with \cite{kocevski26}. This suggests differential evolution in LRDs and non-LRDs even before considering the mass functions. Furthermore, this has key implications for the viable physical models of LRDs.

Most models of LRDs invoke a dense gas envelope around a central engine \citep[e.g.][]{naidu25}, and when examined with medium or higher resolution spectroscopy ($R\gtrsim1000$), most LRDs show blue-shifted absorption in their Balmer lines, potentially indicative of an outflow along the line of sight \citep[e.g.][]{matthee24,kocevski25}. This evidence points to evolution of this envelope with time, likely leading to the ultimate removal or fragmentation of the envelope as it is heated and driven away from the central engine or accreted by the engine. As such, unlike AGN that may experience periods of quiescence between accretion episodes, it may be unlikely for individual LRDs to persist across cosmic time as their proposed gas envelopes are unlikely to remain in place unaltered if the central engine of the LRD shuts down or blows the envelope away. 

Under our fiducial assumption that LRDs are powered by AGN, further AGN emission after the removal of this envelope would likely present as a standard broad-line AGN \citep{jeon26}. \cite{inayoshi25b} supports this model, suggesting that LRDs may be restricted to being the first burst of AGN activity in a black hole. If true, this could serve to explain the much higher LRD fraction seen in broad-line objects at $z\sim8$. Firstly, if the LRD state requires a pristine envelope structure and can only correspond to the first burst of AGN activity, such first bursts and pristine conditions are inherently more common at higher redshift. Secondly, LRDs at $z\sim8$ that destroy their envelopes could present as---and in fact become---non-LRD AGN at lower redshifts, both lowering the counts of LRDs and raising the counts of non-LRDs, thereby driving the transition from an $100\%$ LRD fraction at \mbh $>10^7\,M_{\odot}$ down to $48\%$ by $z\sim5$. This also supports the QS framework put forth by \cite{begelman26}, as the QS structure would be inherently unstable and would evolve into a traditional AGN on a timescale of $\sim$100~Myr.

This narrative may be further supported by \bhmf{}s. Unlike the strong evolution with redshift from non-existence to dominance exhibited by the non-LRDs in Figure~\ref{fig:bhmf_lrd}, the LRDs show only moderate evolution ($\sim$0.5~dex), possibly implying that they are both being created and being removed or evolving into something else with time. As such, it seems likely that the LRD state is a transient phase of black hole growth rather than a distinct population, and may give way to standard broad-line AGN as the surrounding envelope is disrupted.

\begin{figure*}[t]
\includegraphics[width=\textwidth]{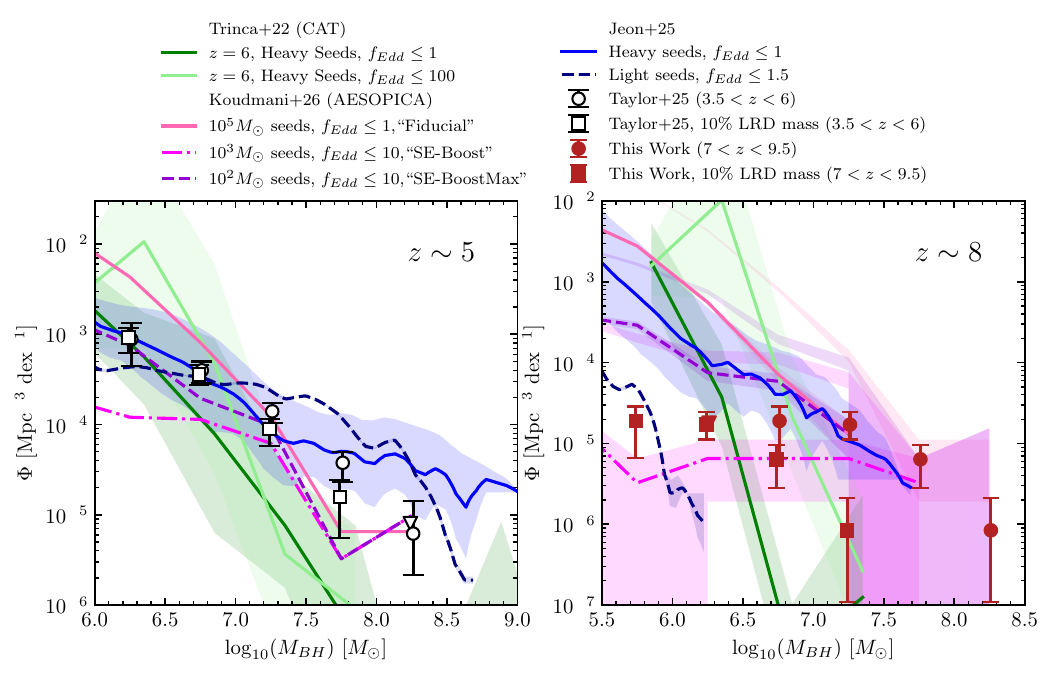}
\caption{This work's observed \bhmf{}s compared to literature models from \cite{Trinca2022}, \cite{jeon25}, and \cite{koudmani26} at $z\sim5$ (left panel) and $z\sim8$ (right panel). Here we plot the models using $10^2$~$M_{\odot}$, $10^3$~$M_{\odot}$, and $10^5$~$M_{\odot}$ seeds from \cite{koudmani26}, models from \cite{jeon25} at $5<z<6$ and $7<z<10$ that use heavy seeds limited to Eddington accretion rates and light seeds permitted to accrete at super-Eddington rates, and models from \cite{Trinca2022} that use heavy seeds with both an Eddington limited model, and a super-Eddington-permitted model. 
Eddington limited light seeds curves for all three simulations do not produce black holes with \mbh{}$>10^6\,M_{\odot}$, and are thus excluded from these figures. This suggests that heavy seeding (solid curves) and/or highly super-Eddington accretion is necessary and prevalent at high redshifts to match observations, regardless of the treatment of LRD mass. }
\label{fig:bhmf_sim}
\end{figure*}

\subsection{Comparison with Empirical and Theoretical Models}\label{sec:models}

In parallel to new observations of high-redshift black hole growth, simulations and models of black hole growth have seen rapid recent advances in the \jwst{} era \citep[e.g.,][]{bhowmick26,dayal26,trinca26}. We now compare our updated \bhmf{}s with literature models from \cite{koudmani26}, \cite{jeon25}, and \cite{Trinca2022} in Figure~\ref{fig:bhmf_sim} to determine the black hole seeding and accretion modes necessary to match our observations.

Briefly, the \cite{koudmani26} models are derived from the \textsc{aesopica} project: a suite of mid-volume ($L = 60$~Mpc) cosmological hydrodynamic simulations based on the \textsc{fable} model \citep{henden18,henden19,henden20}. In Figure~\ref{fig:bhmf_sim}, we show the \cite{koudmani26} models using $10^5$~$M_{\odot}$ black hole seeds limited to Eddington accretion (the ``Fiducial'' simulation in \citealt{koudmani26}), $10^3$~$M_{\odot}$ black hole seeds that may accrete at super-Eddington rates (``SE-Boost''), and $10^2$~$M_{\odot}$ black hole seeds that tend to accrete more often at super-Eddington rates (``SE-BoostMax''). These models exhibit broad agreement with our observed \bhmf{}s only when using heavy seeds, or super-Eddington light seeds. When considering the 10\%-LRD mass \bhmf{}s (square markers in Figure~\ref{fig:bhmf_sim}), these models exhibit a higher normalization \bhmf{} than the observed data, suggesting that intermediate ($10^3$~$M_{\odot}$) seeds, with less extreme super-Eddington accretion could match the observed results. None of the $\leq10^3$~$M_{\odot}$ Eddington limited \cite{koudmani26} models produce enough \mbh{}$\geq10^{5.5}\,M_{\odot}$ black holes at $z\sim8$ to appear on Figure~\ref{fig:bhmf_sim} or match the observed results. 

The results of \cite{jeon25}---which are based on a semi-analytic model with black hole seeding and growth implementations---further drive a need for heavy seeds, as while their sub-Eddington heavy seed model offers a good match to the observed \bhmf{}, even their super-Eddington light seed model fails to match the observed \bhmf{}'s mass range or normalization. As in the \cite{koudmani26} models, adopting the reduced 10\%-LRD masses would permit lighter seeds with less extreme super-Eddington accretion.

Finally, the Cosmic Archaeology Tool (CAT) semi-analytic model \citep{Trinca2022} simulates light (Pop III) and heavy (direct collapse; $M_{\rm seed} \sim 10^5\,M_{\odot}$) seeds in both an Eddington-limited accretion run, and a separate run that permits super-Eddington accretion. As in the \cite{jeon25} models, the CAT light seeds fail to accrete enough mass (even with super-Eddington accretion) to approach the observed \bhmf{} at $z\sim8$, and even the heavy seed models seem to disagree with the normalization and slope of the observed \bhmf{}. However, the CAT models show stronger agreement with the $z\sim8$ \bhmf{} in which the LRD masses are decreased by 90\%. 

Regardless of the treatment of LRD masses, none of the models that use Eddington-limited light seeds produce black holes with \mbh{}$>10^6\,M_{\odot}$ (and are thus not visible in Figure~\ref{fig:bhmf_sim}). While the treatment of super-Eddington accretion varies in each model, only the ``SE-BoostMax'' model from \cite{koudmani26}, in which super-Eddington accretion is most permitted, manages to grow light seeds quickly enough to match observations. Outside of this model, the necessity of heavy seeds in all models to match the \bhmf{} at $z\sim8$, strongly suggests a heavy seed (or highly super-Eddington light--intermediate mass seed) origin for black holes at $z\sim8$. While a 90\% mass reduction for LRDs alleviates some of the need for the heaviest seeds, Eddington-limited light seeds are still insufficient to match the observed data.

\section{Summary}\label{sec:summary}

In this work we present the results of a systematic search for broad-\Hb{} lines at $z>7$ in \jwst{} NIRSpec PRISM spectra. We summarize our main results below. 
\begin{itemize}
\item Starting from a sample of 21,770 uniformly reduced PRISM spectra, we identify 546 objects at $7<z<9.5$ using a bespoke spectroscopic redshift fitting code and visual inspection. We detect broad-\Hb{} emission in 20 unique sources whose broad-\Hb{} emission is not contaminated by outflows. Of these 20 sources, 13 have not been previously published as broad-\Hb{} emitters or interpreted as BLAGN. 
\item We estimate black hole masses for our sample, and after correcting for incompleteness both from the likelihood of successfully detecting broad-lines in PRISM data (primarily a flux-limited effect) and the fraction of sources in our targeted fields that do not yet have NIRSpec observations, we construct the \bhmf{} at $z\sim8$.
\item We compare this \bhmf{} to a reprocessed $z\sim5$ \bhmf{} from \cite{taylor25a} and find the evolution between the two mass functions is easily achievable with modest $f_{\textrm{Edd}}\approx0.1$ black hole growth between the two epochs. 
\item We compare the \bhmf{} to models from \cite{Trinca2022}, \cite{jeon25}, and \cite{koudmani26}, and find that at $z\sim8$, heavy seeds or strongly super-Eddington light seeds are necessary to produce the observed \bhmf, and models with light seeds which grow at the Eddington-limited rate fall far short of matching our observations.
\item We explore two alternative scenarios, one in which the LRDs are not AGN at all, and the other in which the LRD black hole masses are overestimated by a factor of ten. In the first case, our entire $z\sim8$ sample is eliminated, suggesting that non-LRD BLAGN must appear very quickly between $z=8-5$ in order to produce the abundance of BLAGN observed at $z\sim5$, or that LRDs may evolve into non-LRD BLAGN. In the second scenario, where we assume that the LRD masses are overestimated by a factor of ten, we find that the normalization of the \bhmf{} would be moderately reduced, but heavy seeds or strongly super-Eddington light seeds are still necessary for the models to match the data.  
\item We estimate bolometric luminosities for the BLAGN sample using the \cite{greene26} scaling factors and compute a BHAD of $4.8^{+5.8}_{-2.6}\times10^{-7}\,M_{\odot}\,\textrm{yr}^{-1}\,\textrm{Mpc}^{-3}$ using a Schechter fit to the resulting \lbol{} luminosity function or a lower limit of BHAD$=2.3^{+0.6}_{-0.3}\times10^{-7}~M_{\odot}~\textrm{yr}^{-1}~\textrm{Mpc}^{-3}$ when simply summing the accretion of the BLAGN sample. These values show broad agreement with previous photometric \jwst{} studies of AGN candidates and LRDs at $z>5$ \citep{akins25c,umeda26}, as well as the model from \cite{jeon22}, but are inconsistent with an extrapolation of $z<5$ X-ray studies \citep{aird15}. We further derive a completeness-corrected mean $f_{\textrm{Edd}}\approx0.08$ for the sample (assuming that the LRD masses are consistent with virial relations), showing broad agreement with the \bhmf{} evolution estimate above.
\item We measure the UV luminosity function for the BLAGN sample, and---as at $z\sim5$---find a $\sim1-2$~dex excess in the UV luminosity function normalization for AGN when compared to pre-\jwst{} studies. 
\end{itemize}

Overall, this sample of 20 broad-\Hb{} emitters at $7<z<9.5$ offers the first insights into LRD and/or BLAGN populations in the first $\sim$750~Myr of the universe, getting us ever closer to the initial SMBH formation epoch. As in previous studies of BLAGN at $z\sim5$, the presence of this sample defies pre-\jwst{} expectations, indicating that BLAGN grew far larger and far faster than expected. While pushing to even higher redshifts ($z>10$)---where the \Hb{} line will no longer be observable with \jwst{}/NIRSpec---will be challenging, this sample only scratches the surface of BLAGN studies at $7<z<9.5$. Deeper medium or high-resolution grating data, or IFU studies of this sample could greatly enhance our understanding of the broad \Hb{} line structure, and thus the physical conditions present in these objects. MIRI observations of broad-H$\alpha$ may offer a way forward, and NIRSpec/PRISM followup of the many remaining $7<z<9.5$ galaxy candidates in both lensed and blank fields will expand the known population of BLAGN. The summation of these efforts would allow for enhanced number statistics in the \bhmf{} and increase the likelihood of discovering the rarest most extreme BLAGN in the early universe. They in turn are unique probes of structure formation at the (small-scale and high-redshift) tail end of $\Lambda$CDM cosmology. 

\begin{acknowledgements}
AJT acknowledges support from the UT Austin College of Natural Sciences and STScI through award \jwst{}-GO-6368. PGP-G acknowledges support from grants PID2022-139567NB-I00 and PID2025-169195NB-I00 funded by Spanish Ministerio de Ciencia, Innovaci\'on y Universidades MICIU/AEI/10.13039/501100011033, and the European Union FEDER program {\it Una manera de hacer Europa}. 
PAH acknowledges support from NASA under award 80GSFC24M0006.
SK acknowledges support from the Royal Society under grant number URF\textbackslash R1\textbackslash 251867.

\end{acknowledgements}

\software{astropy: \citep{astropy:2013,astropy:2018,astropy22}, \texttt{BAGPIPES}: \citep{carnall18}, emcee: \citep{emcee}, \texttt{grizli}: \citep{grizli}, \texttt{lazy.jl}: \citep{lazy_jl}, \texttt{msaexp}: \citep{msaexp}, \texttt{pandiea\_engine}: \citep{pontoppidan16}, scipy: \citep{scipy20}, SpectRes: \citep{spectres}}
\facilities{\jwst{}}

\bibliographystyle{aasjournalv7.1}
\bibliography{latest.bib}

\newpage
\begin{appendix}

\section{Supplementary Tables}\label{apdx:table}

\begin{deluxetable*}{rrrccccc}[h]
\tablecaption{Spectra Contributing to the BLAGN Sample} \label{tab:apdx}
\tablehead{Source ID & R.A. & Decl. & $z_{spec}$ & Exposure Time & Field & Reference & Spectra \\ 
 & [deg] & [deg] & & [hours]}
\startdata
J001419.16--302405.6 & 3.579829 & -30.401569 & 7.040 & 21.01 & A2744 & [1,2] & 2561-2-13123, 2561-6-24175 \\
J100003.90+020734.2 & 150.016232 & 2.126173 & 7.060 & 1.46 & COSMOS & New & 7076-11-114105 \\
J021659.48--050907.2 & 34.247828 & -5.151996 & 7.100 & 5.17 & UDS & New & 4233-2-138196, 5224-2-920396 \\
J141943.08+525316.5 & 214.929512 & 52.887913 & 7.155 & 6.99 & CEERS & New & 4233-6-42803, 6368-65-14197, 7076-20-15706 \\
J021738.29--051919.4 & 34.409532 & -5.322046 & 7.180 & 1.46 & UDS & New & 7076-18-73433 \\
J021754.78--051712.6 & 34.478255 & -5.286832 & 7.210 & 0.79 & UDS & New & 4233-3-20030 \\
J021729.37--050657.3 & 34.372370 & -5.115912 & 7.220 & 1.58 & UDS & New & 4233-2-163748, 4233-3-163748 \\
J033228.58--275333.2 & 53.119075 & -27.892556 & 7.408 & 2.31 & GOODS-S & New & 1287-3-11786 \\
J095918.44+021420.4 & 149.826836 & 2.239006 & 7.510 & 1.06 & COSMOS & New & 7417-14-22794 \\
J100138.48+021912.1 & 150.410345 & 2.320026 & 7.620 & 1.46 & COSMOS & New & 7076-2-786768 \\
J095948.35+022655.5 & 149.951441 & 2.448757 & 7.650 & 1.46 & COSMOS & [3] & 7076-9-853717 \\
J021735.40--050806.4 & 34.397484 & -5.135116 & 7.760 & 4.38 & UDS & [4] & 5224-1-150135 \\
J100120.74+021438.8 & 150.336418 & 2.244101 & 8.140 & 1.46 & COSMOS & New & 7076-10-1068405 \\
J095920.97+021329.8 & 149.837357 & 2.224933 & 8.300 & 1.06 & COSMOS & New & 7417-14-21076 \\
J021718.46--051326.2 & 34.326908 & -5.223947 & 8.310 & 1.58 & UDS & New & 6368-37-29908 \\
J141930.27+525251.0 & 214.876139 & 52.880827 & 8.351 & 5.53 & CEERS & [5] & 4233-6-966323, 6368-49-8752 \\
J001433.70--302311.2 & 3.640408 & -30.386438 & 8.500 & 2.63 & A2744 & [6] & 2561-2-20466 \\
J114933.82+222059.2 & 177.390929 & 22.349767 & 8.630 & 0.79 & MJ1149 & [7] & 1208-49-5112687 \\
J095940.19+021725.5 & 149.917473 & 2.290408 & 8.960 & 1.46 & COSMOS & New & 7076-5-826738 \\
J100032.70+021828.9 & 150.136253 & 2.308030 & 9.288 & 4.74 & COSMOS & [8] & 6368-10-119334 \\
\enddata
\tablecomments{Spectra are identified as [\jwst{} Program ID]-[Observation Number]-[Source ID]. \newline [1] \cite{greene24}, [2] \cite{furtak24}, [3] \cite{lin26}, [4] \cite{naidu25}, [5] \cite{wang24b}, [6] \cite{kokorev23}, [7] \cite{tripodi25}, [8] \cite{taylor25c}.}
\end{deluxetable*}

\begin{deluxetable*}{cccccc}[h]
\tablecaption{$3.5<z<6$ Black Hole Mass Function (all objects)} \label{tab:bhmfz5all}
\tablehead{$\log\left(M_{BH}/M_{\odot}\right)$ & $\Phi$ [$\times10^{-6}$] & N & \multicolumn{3}{c}{Completeness} \\
 & Mpc$^{-3}$\,dex$^{-1}$ & & (total) & (obs) & (det)}
\startdata
6.25 & $880^{+447}_{-430}$ & 5.56 & 0.01 & 0.03 & 0.17 \\
6.75 & $399^{+101}_{-113}$ & 16.83 & 0.03 & 0.13 & 0.29 \\
7.25 & $140^{+34.1}_{-35.4}$ & 19.65 & 0.11 & 0.21 & 0.54 \\
7.75 & $37.8^{+12.3}_{-14.6}$ & 9.88 & 0.21 & 0.25 & 0.86 \\
8.25 & $6.20^{+7.95}_{-4.04}$ & 2.03 & 0.26 & 0.26 & 0.99 \\
\enddata
\tablecomments{Reprocessed from \cite{taylor25a}. ``det'' refers to line detection completeness, ``obs'' refers to observed fraction completeness, and ``total'' is the product of  ``det'' and ``obs''.}
\end{deluxetable*}

\begin{deluxetable*}{cccccc}[h]
\tablecaption{$3.5<z<6$ Black Hole Mass Function (non-LRDs)} \label{tab:bhmfz5agn}
\tablehead{$\log\left(M_{BH}/M_{\odot}\right)$ & $\Phi$ [$\times10^{-6}$] & N & \multicolumn{3}{c}{Completeness} \\
 & Mpc$^{-3}$\,dex$^{-1}$ & & (total) & (obs) & (det)}
\startdata
6.25 & $871^{+571}_{-382}$ & 5.06 & 0.01 & 0.03 & 0.17 \\
6.75 & $336^{+113}_{-97}$ & 13.26 & 0.03 & 0.12 & 0.29 \\
7.25 & $87.1^{+28.2}_{-31.4}$ & 10.78 & 0.10 & 0.20 & 0.50 \\
7.75 & $15.8^{+8.3}_{-10.3}$ & 3.89 & 0.19 & 0.23 & 0.84 \\
8.25 & $<8$ & 0.98 & 0.19 & 0.20 & 0.99 \\
\enddata
\tablecomments{Reprocessed from \cite{taylor25a}. ``det'' refers to line detection completeness, ``obs'' refers to observed fraction completeness, and ``total'' is the product of  ``det'' and ``obs''.}
\end{deluxetable*}

\begin{deluxetable*}{cccccc}[h]
\tablecaption{$3.5<z<6$ Black Hole Mass Function (LRDs)} \label{tab:bhmfz5lrd}
\tablehead{$\log\left(M_{BH}/M_{\odot}\right)$ & $\Phi$ [$\times10^{-6}$] & N & \multicolumn{3}{c}{Completeness} \\
 & Mpc$^{-3}$\,dex$^{-1}$ & & (total) & (obs) & (det)}
\startdata
6.25 & $<18.8$ & 0.50 & 0.04 & 0.15 & 0.26 \\
6.75 & $62.8^{+41.3}_{-38.8}$ & 3.57 & 0.05 & 0.15 & 0.29 \\
7.25 & $53.0^{+18.3}_{-21.8}$ & 8.88 & 0.13 & 0.23 & 0.59 \\
7.75 & $21.9^{+8.7}_{-11.5}$ & 5.99 & 0.22 & 0.25 & 0.87 \\
8.25 & $2.20^{+4.69}_{-1.84}$ & 1.05 & 0.38 & 0.38 & 1.00 \\
\enddata
\tablecomments{Reprocessed from \cite{taylor25a}. ``det'' refers to line detection completeness, ``obs'' refers to observed fraction completeness, and ``total'' is the product of  ``det'' and ``obs''.}
\end{deluxetable*}

\end{appendix}

\end{document}